\documentclass[paper]{JHEP3}
\usepackage{graphics}
\usepackage{amsmath}
\usepackage[dvips]{graphicx}
\usepackage{amsfonts,amsbsy}
\usepackage{amssymb}
\usepackage{bbm}
\newcommand{\beq}{\begin{equation}}
\newcommand{\eeq}{\end{equation}}
\newcommand{\beqa}{\begin{eqnarray}}
\newcommand{\eeqa}{\end{eqnarray}}
\def\r{{\boldsymbol r}}
\def\z{{\boldsymbol z}}
\def\x{{\boldsymbol x}}

\def\k{{\boldsymbol k}}
\def\q{{\boldsymbol q}}
\def\p{{\boldsymbol p}}
\def\0{{\boldsymbol 0}}

\def\cal{\mathcal}

\def\bra#1{\langle#1\vert}
\def\ket#1{\vert#1\rangle}

\def\simle{\mathrel{\rlap{\raise 0.511ex \hbox{$<$}}{\lower 0.511ex 
\hbox{$\sim$}}}}
\def\simge{\mathrel{ \rlap{\raise 0.511ex 
\hbox{$>$}}{\lower 0.511ex \hbox{$\sim$}}}}

\title{A path integral for heavy-quarks in a hot plasma}

\author{A. Beraudo$^{1,2,3}$\\
$^1$Centro Studi e Ricerche \emph{Enrico Fermi},\\
Comprensorio del Viminale, Piazza del Viminale 1, Roma, ITALY;\\
$^2$Physics Department, Theory Unit, CERN, CH-1211 Gen\`eve 23, Switzerland;
$^3$Dipartimento di Fisica Teorica dell'Universit\`a di Torino and\\ Istituto Nazionale di Fisica Nucleare, Sezione di Torino,\\
Via Pietro Giuria 1, 10154 Torino, ITALY\\
E-mail: \email{beraudo@to.infn.it}}
\author{J.P. Blaizot\\
Institut de Physique Theorique, CEA-Saclay,
91191 Gif-sur-Yvette\\
E-mail: \email{blaizot@cea.fr}}
\author{P. Faccioli$^{4,5}$\\
$^4$Dipartimento di Fisica dell'Universit\`a di Trento and \\
$^5$Istituto Nazionale di Fisica Nucleare, Sezione di Trento (Padova)\\
Via Sommarive 14, I-38100 Povo, Trento\\
E-mail: \email{faccioli@science.unitn.it}}
\author{G. Garberoglio$^{4,6,7}$\\
$^6$CNISM, Consorzio Nazionale Interuniversitario per le Scienze Fisiche della
Materia;\\
$^7$Interdisciplinary Laboratory for Computational Science (LISC), FBK-CMM and
University of Trento,\\
via Sommarive 18, I-38123 Povo, Trento\\ 
E-mail: \email{garberoglio@fbk.eu}}

\abstract{We propose a model for the propagation of a heavy-quark in a hot plasma, to be viewed as a first step towards a full description of the dynamics of heavy quark systems in a quark-gluon plasma, including bound state formation. The heavy quark is treated as a non relativistic particle interacting with a fluctuating field, whose correlator is determined by a hard thermal loop approximation. This approximation, which concerns only the medium in which the heavy quark propagates, is the only one that is made, and it can be improved. The dynamics of the heavy quark is given exactly by a quantum mechanical path integral that is calculated in this paper in the Euclidean space-time using numerical Monte Carlo  techniques. The spectral function of the heavy quark in the medium is then reconstructed using a Maximum Entropy Method. The path integral is also evaluated exactly in the case where the mass of the heavy quark is infinite; one then recovers known results concerning the complex optical potential that controls the long time behavior of the heavy quark. The heavy quark correlator and its spectral function is also calculated semi-analytically at the one-loop order, which allows for a detailed description of the coupling between the heavy quark and the plasma collective modes.
}


\begin{document}
\maketitle
\section{Introduction}
Understanding the dynamics of heavy quarks in a quark-gluon plasma, and the fate of their possible bound states, has remained a difficult issue, ever since the original proposal of Matsui and Satz \cite{Mat} to view the dissolution of  $J/\psi$'s mesons in a quark-gluon plasma  as a signal of deconfinement (for a recent review see for instance Ref.~\cite{david}).  Aside from many studies based on the assumption that the dominant effect of the plasma is to screen the interaction potential, more recently, the problem has been attacked using a ``first principle'' approach, namely by calculating the $Q\overline{Q}$ spectral functions reconstructed from the corresponding Euclidean correlators provided by lattice QCD. The melting of the $J/\psi$, for instance, is then signaled by the disappearance of the corresponding peak in its  spectral function. 
The first results of such an analysis led to the surprising result that the $J/\psi$ appears to survive till temperatures well above $T_c$~\cite{dat,asa1,asa2,aarts,petre1}, in sharp contrast with studies based on screened potentials. A comparison between correlators and spectral functions evaluated on the lattice and within different potential models was attempted in~\cite{mocsy,albe2,rapp}, revealing ambiguities in the whole procedure. Another line of first principle calculations  was undertaken in a number of recent papers~\cite{lai1,lai2,bbr,bra,vai,du}: in these works, the correlator of a heavy quark pair is calculated directly in real time, revealing that the long time behavior of the dynamics can be encompassed by a Schr\"odinger equation with a complex potential that describes both the effects of screening and, through its imaginary part, of the collisions with the plasma particles. 

While it represents an important step forward, this description of the dynamics of heavy quarks by a Schr\"odinger equation and an effective potential has  limitations. The potential is calculated, and well defined, only in the limit of infinitely massive quarks. Moreover, a simple potential description emerges only at large times, that is, at time scales that are large compared to the typical times characterizing the response of the plasma to perturbations. In the situations which we want eventually to deal with,  namely the fate of bound states of heavy quarks in the environment created in ultra-relativistic heavy ion collisions, all relevant time scales are mixed  (see for instance \cite{BO82}), and a description of the dynamics beyond that provided by a simple Schr\"odinger equation is called for. This paper represents an attempt in this direction, building on the approach developed in \cite{bbr}. Our strategy, already sketched in \cite{QM}, is the following. The heavy quarks are treated as massive, non relativistic,  particles coupled to a fluctuating gauge field. 
The dynamics of the heavy quark is then encoded  exactly in a quantum mechanical path integral, while the average over the gauge field fluctuations is entirely determined by the properties of the medium. If one restricts oneselves to  approximations where this average is Gaussian, and hence can be performed analytically, the gauge fields can be eliminated completely, leaving a path integral for a non relativistic particle with a non local (in space and time) self-interaction term. This path integral is reminiscent of that introduced by Feynman in his treatment of the ``polaron''  \cite{Feynman}. An approximation that leads to a Gaussian average (at least in the Abelian case), is the hard thermal loop approximation (HTL) \cite{HTL}. We shall make use of such an approximation, because of its simplicity, and also because it encompasses the dominant plasma effects that one wants to include: screening effects, collective modes, and collisions.  We emphasize, however, that this approximation, which concerns primarily the medium in which the heavy quark propagates,  can be improved without altering the basic structure of the problem.

The present paper has an exploratory character and represents only a first step in this long-term goal. 
 It focusses on the dynamics of a single heavy quark moving in a plasma of light charged particles, for which we provide a simple model. We use for the quark-gluon plasma an idealization where only Abelian (in fact, Coulomb) interactions are taken into account. We also assume, for simplicity, that the plasma particles are fermions, i.e., quarks. In short, we model the quark-gluon plasma by an electromagnetic plasma, treated within the HTL approximation. This is enough to take into account typical plasma effects, such as screening, Landau damping of collective excitations, and collisions between the heavy quark and the plasma particles. These phenomena are characterized by a single scale, the Debye screening mass $m_D$,  to which, in our numerical studies, we  shall give a value characteristic of a quark-gluon plasma at a given temperature (thereby taking effectively gluons into account).  
The dynamics of the heavy quarks is then treated exactly within a path integral of the type discussed above, with a non-local self-interaction whose space-time properties are controlled by the Debye mass.

Our paper is organized as follows. In Sect.~\ref{sec:general} we establish the general setting:
the basic properties of the propagator of a heavy particle are recalled, a description of the medium of light charged particles in which the propagation takes place is given, the path integral for the heavy quark propagator is constructed. This path integral is calculated exactly in the limit of an infinitely massive quark, and known results are recovered concerning the long time behavior of the heavy quark propagator in this limit. Then, in 
Sect.~\ref {sec:physical_processes}, we calculate the heavy quark propagator in the one-loop approximation, providing a detailed analysis of the coupling of the heavy quark to the collective plasma excitations and of  the role of the collisions. We also calculate the spectral function and the resulting Euclidean correlator. In Sect.~\ref{sec:numerical} we present the results of the numerical evaluation of  the path integral in Euclidean space-time, using Monte Carlo (MC) techniques. We use the Maximum Entropy Method (MEM) to  reconstruct the spectral density. Within our present implementation of this method, we can only reproduce, semi-quantitatively, the main features of the spectral density. Finally,  Sect.~\ref{sec:conclusions} summarizes the conclusions. 
In Appendix~\ref{sec:toy-model} we give a self-contained presentation of an exactly solvable  toy-model which captures general features of the heavy quark correlator and its spectral function, and this for any value of the coupling constant.\footnote{A somewhat similar model, with however different emphasis, was considered in Ref.~\cite{young}.}
Appendix \ref{sec:details} provides details on the numerical
evaluation of the path integral.

\section{A path integral for the heavy-quark propagator}\label{sec:general}

In this section, we recall general properties of the heavy quark propagator, and establish the basic path integral that describes the dynamics of the heavy quark coupled to a gauge field that is integrated out via a Gaussian averaging.

\subsection{The heavy quark-propagator. Generalities}
\label{sec:generalities}

Most of the physical information that we are interested in can be obtained from the study of the following correlator
\beq\label{eq:single>}
G^>(t,\r|0,{\bf 0})\equiv\langle\psi(t,\r)\psi^\dagger(0,{\bf 0})\rangle,
\eeq
where $\psi(t,\r)$ denotes the heavy quark field. In the following we  shall most of the time use the simplified notation $G^>(t,\r)$ for $G^>(t,\r|0,{\bf 0})$. The expectation value in the above formula  is a thermal average over the states of  light particles (with no heavy quark present) that will be specified later. 
At this stage, we simply note that the full  Hamiltonian $H$  can  be decomposed into three contributions:  
\beq\label{eq:hamiltoniandecomp}
H=H_Q+H_{med}+H_{int},
\eeq
where $H_Q$ is the (non relativistic) Hamiltonian describing the heavy quark in vacuum, 
$
H_{med}$ is the Hamiltonian of the medium in which the heavy quark propagates, and 
$
H_{int}$
represents the interactions between the medium and the heavy quarks.  
For the parts that depend explicitly on the fermion field, we have
\beq
H_Q=M\int d^3 \r\, \psi^\dagger(\r)\psi(\r)+\int d^3 \r \,\psi^\dagger(\r)\left( -\frac{\nabla^2}{2M} \right)\psi(\r),
\eeq
and 
\beq
H_{int}=g \int d^3\, \r \,\psi^\dagger(\r)\psi(\r) A_0(\r),
\eeq
where $A_0(\r) $ represents the local electrostatic potential created by the light particles. 
The full Hamiltonian commutes with the number of heavy quarks $N_Q$: 
\beq\label{commute}
[H,N_Q]=0,\qquad N_Q=\int d^3\r\,  \psi^\dagger(\r)\psi(\r),
\eeq
and one can classify its eigenstates according to the number of heavy quarks that they contain. In particular, one may write a spectral decomposition of the correlator (\ref{eq:single>}):
\beq\label{spectraldecomp}
G^>(t,\r)= \sum_{n,\bar m}\frac{{\rm e}^{-\beta E_n}}{Z} {\rm e}^{i(E_n-E_{\bar m})t}\bra{n} \psi(\r)\ket{\bar m}\bra{\bar m}\psi^\dagger({\bf 0}) \ket{n},
\eeq
where the states $\ket{n}$ contain no heavy quark, while the states $\ket{\bar m}$ contain one heavy quark, i.e., 
\beq
\psi(\r)\ket{n}=0,\qquad N_Q\ket{\bar m}=\ket{\bar m}.
\eeq
In eq.~(\ref{spectraldecomp}) $Z$ is the partition function of the system without heavy quark. 
It follows also from Eq.~(\ref{commute}) that $G^<(t,\r)\equiv\langle \psi^\dagger(0,0)\psi(t,\r)\rangle=0$, so that  the time-ordered propagator, 
$
G(t)\equiv i\,\langle {\rm T} \psi(t)\psi^\dagger(0)\rangle= i\,\theta(t)G^>(t)-i\,\theta(-t)G^<(t),
$
and  the retarded propagator
$
G_R(t)\equiv i\,\theta(t)\left[G^>(t)+G^<(t)\right]
$
are identical, $G(t)=G_R(t)= i\,\theta(t)\, G^>(t).$
Similarly, the spectral density is given  by the Fourier transform of Eq.~(\ref{eq:single>}), namely:
\beq\label{eq:HQspec}
\sigma(\omega)\equiv G^>(\omega)+G^<(\omega)=G^>(\omega)=\int_{-\infty}^{\infty} dt \,{\rm e}^{i\omega t} \,G^>(t),
\eeq
with the inverse relation
\beq\label{euclcorrelspec}
G^>(-i\tau)=\int_{-\infty}^{+\infty}\frac{d\omega}{2\pi}e^{-\omega\tau}\sigma(\omega).
\eeq
In this last equation we have exploited the analyticity of $G^>(t)$ in the strip $-\beta<{\rm Im} \, t<0$, and set $t=-i\tau$, with $0<\tau<\beta$. Inverting this relation, namely calculating $\sigma(\omega)$ from the Euclidean correlator $G^>(-i\tau)$ is a difficult (well known) problem that we shall address briefly in the last part of this paper.

By noting that $H_{med}$ does not depend on $\psi$, one finds (with all fields in the Heisenberg representation)
\beq
[\psi,H]=[\psi, H_Q+H_{int}]=\left(M-\frac{\nabla^2}{2M}+gA_0(t,\r)  \right)\psi(t,\r),
\eeq
so that, from the equation of motion 
$i\partial_t \psi(t,\r)=[\psi,H],$ we get
\beq\label{eqmG>}
i\partial_t G^>(t,\r)=\left(M-\frac{\nabla^2}{2M}  \right)G^>(t,\r)+g\langle A_0(t,\r)\psi(t,\r)\psi^\dagger({\bf 0})\rangle.
\eeq
In the absence of interactions, this equation has the familiar solution 
\beq\label{freesolution}
G_0^>(t,\r)={\rm e}^{-iMt}\left(  \frac{M}{2\pi i t} \right)^{3/2}{\rm e}^{i M\r^2/2t}, 
\eeq
corresponding to the initial condition $G_0^>(t=0,\r)=\delta(\r)$. Note that this initial condition still holds in the presence of interactions, i.e., $G^>(t=0,\r)=\delta(\r)$, as is easily verified. A further exact relation is obtained by considering the equation (\ref{eqmG>}) at $t=0$:
\beq\label{der1t0}
i\left. \partial_t G^>(t,\r)\right|_{t=0}=\left(M-\frac{\nabla^2}{2M}+g\langle A_0(t=0,\r)\rangle  \right)\delta(\r).
\eeq
Since the thermal average involves only states of the medium which do not contain heavy quarks that could polarize it, we have $\langle A_0(\r)\rangle=(1/Z) \sum_n {\rm e}^{-\beta E_n} \bra{n} A_0(\r)\ket{n}=0$: the interactions do not contribute to the leading (linear) order in a small time expansion. Pushing this expansion to second order, one gets
\beq\label{der2t0}
-\partial_t^2 \left. G^>(t,\r)\right|_{t=0}=\left[ \left(M-\frac{\nabla^2}{2M}  \right)^2+g^2\langle A_0^2(0,\r)\rangle \right]\delta(\r), 
\eeq
or, taking a Fourier transform
\beq\label{der2t0FT}
-\partial_t^2 \left. G^>(t,\p)\right|_{t=0}=\left(M+\frac{\p^2}{2M}  \right)^2+g^2\langle A_0^2\rangle ,
\eeq
where $\langle A_0^2\rangle$ stands for $\langle A_0^2(t=0,\r=0)\rangle$.
Thus at order $t^2$, the effect of the interaction is governed by the size of the fluctuations of $A_0$, an intrinsic property of the medium to be discussed further later. Note also that the coefficient of $t^2$ in the expansion of  $G^>(t,\p)$ at small $t$ is of order $g^2$. 

We may turn these relations for the derivatives of $G^>(t,\p)$ at $t=0$ into sum rules for the spectral density. From the initial condition (see after Eq.~(\ref{freesolution})), and the relations (\ref{der1t0}) and  (\ref{der2t0}) above, one gets, respectively,
\beqa\label{sumrules012}
\int_{-\infty}^{\infty} \frac{d\omega}{2\pi} \sigma(\omega,\p)=1,\quad \int_{-\infty}^{\infty} \frac{d\omega}{2\pi} \,\omega\sigma(\omega,\p)=M+\frac{\p^2}{2M},\qquad \qquad \qquad\nonumber\\   \int_{-\infty}^{\infty} \frac{d\omega}{2\pi} \,\omega^2\sigma(\omega,\p)=\left(M+\frac{\p^2}{2M}\right)^2+g^2\langle A_0^2\rangle.
\eeqa
The last sum rule assumes that $\langle A_0^2\rangle$ is well defined. However, as we shall see in the next subsection,  within the approximation used in the present paper $\langle A_0^2\rangle$ is in fact given by a divergent integral, so that this sum rule will not apply. Accordingly the short time behavior of the correlator will not have a simple Taylor expansion as assumed in the discussion above (Eq.~(\ref{der2t0FT})). 

The Fourier transform of $G^>(t,\r)$ used above (see Eq.~(\ref{der2t0FT})) is of the form\beq
G^>(t,\p)=\int d^3r\, {\rm e}^{-i\p\cdot\r} G^>(t,\r)=\langle a_\p(t) a_\p^\dagger\rangle, 
\eeq
where $a_\p$ and $a_\p^\dagger$ are the Fourier transform of the field operators $\psi(\r)$ and $\psi^\dagger  (\r)$, respectively, and we have used the translation invariance of the medium in order to implement the conservation of the total momentum  ($\langle a_\p a^\dagger_{\p'}\rangle\sim \delta_{\p,\p'}$). For the value $t=-i\beta$, the correlator $G^>(-i\beta,\p)$ yields the difference of the free energies of the systems with and without a heavy quark. To see that, note that this free energy difference is given by 
\beqa\label{free_energy_p}
\exp[-\beta\Delta F_{Q,\p}]&=&\frac{1}{Z}\sum_n\langle n|a_\p\,e^{-\beta H}\,a_\p^\dagger|n\rangle\nonumber \\
{}&=&\frac{1}{Z}\sum_n e^{-\beta E_n}\langle n|a_\p(\beta)\,a_\p^\dagger(0)|n\rangle=G^>(-i\beta,\p).
\eeqa
In the first line of Eq.~(\ref{free_energy_p}), the states $a^\dagger_\p\ket{n}$, while not  eigenstates of $H$, constitute a basis of states  with momentum $\p$ and containing one heavy quark. Thus,  the sum over the states $\ket{n}$ in the first line of Eq.~(\ref{free_energy_p}) is indeed the partition function for the system with one heavy quark and total momentum $\p$. 

In preparation for the forthcoming discussion, let us recall that the propagator of the heavy quark, treated as a non relativistic quantum mechanical particle, may be given a path integral representation \cite{Feynman}. With  $A_0(t,\x)$ considered as a given external potential, we can write  ($t>0$):
\beq
G^>(t,\r)=\int_{0}^{\r} {\cal D}\z \,\exp\left[   i\int_0^t dt' \left( \frac{1}{2}M\dot\z^2 -g A_0(t,\z)  \right)\right], 
\eeq
where the symbol $\int_{0}^{\r}{\cal D}\z$ indicates a path integration over paths $\z(t)$ such that $\z(0)=0$ and $\z(t)=\r$.
The transcription of this expression in imaginary time reads ($\tau>0$):
\beq
G^>(-i\tau,\r)=\int_{0}^{\r} {\cal D}\z\, \exp\left[   -\int_0^\tau d\tau' \left( \frac{1}{2}M\dot\z^2 +ig A^E_0(\tau,\z)  \right)\right],
\eeq
where, aside from making the familiar substitution $t\to -i\tau$, we have also introduced the Euclidean field $A_0^E(\tau,\r)=-iA_0(t=-i\tau,\r)$.

\subsection{A model for the medium}
\label{sec:model_medium}

The medium is modeled by a plasma of light fermions with Coulomb interactions. Because of its large mass, the heavy quark has a small velocity, and consequently its ability to induce magnetic excitations of the medium is small; accordingly these magnetic excitations will be ignored. 
The Hamiltonian reads then
\beq
H_{med}=\int d^3 r\, \xi^\dagger (\r) h_0\, \xi(\r)+\frac{1}{2}\int d^3r d^3 r' \hat\rho(\r) \frac{g^2}{4\pi|\r-\r'|}\hat\rho(\r'),
\eeq
where $\xi(\r)$ and $\xi^\dagger(\r)$  denote the field operators of the light fermions, $\hat\rho(\r)\equiv \xi^\dagger(\r)\xi(\r)$ is the charge density of the light particles, and $h_0$ their  free Hamiltonian.

To the full Hamiltonian of the system corresponds an Euclidean action of the form $S=S_Q+S_{int}+S_{med}$, with 
\beq
S_{med}=\int_x\, \xi^*(x) (\partial_\tau+h_0)\,\xi(x)+\frac{g^2}{2}\int_{x,x'} \rho(x) K(x,x')\rho(x'),
\eeq
and 
\beq
S_{int}={g^2}\int_{x,x'} \rho_Q(x) K(x,x')\rho(x').
\eeq
We have set $\rho_Q(x)=\psi^*(\tau,\r)\psi(\tau,\r)$, and
\beq
\int_x\equiv\int d^4 x \equiv \int_0^\beta d\tau \int d^3 r, \qquad x\equiv (\tau,\r).
\eeq
The operator $K(x,x')$ is given by 
\beq
- \nabla_\r^2 K(x,x')=\delta(x-x'), \qquad K(x,x')=\delta(\tau-\tau')\,\frac{1}{4\pi|\r-\r'|}.
\eeq
In calculating the partition function of the system, one can proceed in a familiar way,  and  integrate over the light fermions after eliminating their density $\rho(x)$ in favor of a gauge potential $A_0^E(x)$ ($-igK\cdot (\rho+\rho_Q)\rightarrow A_0^E$). One then obtains
\beqa\label{Zpathint1}
\int {\cal D}(\xi^*,\xi)\,{\rm e}^{-\left(S_{int}+S_{med}\right)}=\int {\cal D}A_0^E\, {\rm e}^{-S[A_0^E]}\, ,
\eeqa
where 
\beqa \label{Zpathint2}
S[A_0^E]&=& ig\int_x A_0^E(x)\rho_Q(x)\nonumber\\ &-&{\rm Tr} \ln(\partial_\tau+h_0+igA_0^E)+\frac{1}{2}\int_{x,x'}A_0^E(x)K^{-1}(x,x')A_0^E(x'),
\eeqa
and the field $A_0^E$ obeys periodic boundary conditions in imaginary time, $A_0^E(0,\r)=A_0^E(\beta,\r)$, reflecting the fact that the medium of light particles is in thermal equilibrium at temperature $T=1/\beta$.

At this point we perform the main approximations that will yield a simple model for the medium. In the expansion of the fermionic determinant (the second term in the r.h.s. of  Eq.~(\ref{Zpathint2})) in powers of $A_0^E$, we keep only the quadratic term. Furthermore, we keep only the leading high temperature approximation to the corresponding 2-point function (the so-called hard-thermal-loop (HTL) approximation \cite{HTL}). Note that in the case of QED, the HTL  approximation automatically truncates the expansion of the determinant at quadratic order. Further discussion of the validity of this approximation will be made shortly. At this point we note that once it is done,  we are left with a simple quadratic action:
\beqa\label{Zpathint3}
S[A_0^E]=ig\int_x \,A_0^E(x)\rho_Q(x)+ \frac{1}{2}\int_{x,x'} A_0^E(x)\tilde\Delta^{-1}(x,x')A_0^E(x'). 
\eeqa
The propagator $\tilde\Delta(x,x')=\langle A_0^E(x) A_0^E(x')\rangle$ is given in Fourier space by
$\tilde\Delta^{-1}(z,\q)=q^2+\Pi(z,\q)$, 
where $\Pi(z,\q)$ is the (longitudinal) polarization tensor in the Coulomb gauge:
\beq\label{polar}
\Pi(z,\q)=m_D^2 \left(1-Q(z/q)\right),\qquad  Q(x)\equiv \frac{x}{2}\ln\frac{x+1}{x-1},
\eeq
with $m_D=\Pi(z=0,\q)$ is the Debye mass. The Debye mass is the mass scale that characterizes the response of the medium. 

At this point, we note that the equations we have written hold exactly only for the hot electromagnetic plasma. However, at this level of approximation, the main difference with a quark-gluon plasma lies in the value of the Debye mass that, in a QCD plasma,  receives also contributions from gluons. In the numerical studies to be presented below, in order to get orders of magnitudes that are relevant for the quark-gluon plasma, we shall adjust the Debye mass to the value it would have in a quark-gluon plasma at the considered temperature, that is we shall use the QCD HTL expression $m_D=g_s^2T^2\left({N_c}/{3}+{N_f}/{6}\right)$, with  $g_s^2/4\pi=\alpha_s$ the strong coupling constant. With $\alpha_s=0.3$, $N_c=3$ and $N_f=2$, this yields a value $m_D=713$ MeV for $T=300$ MeV. Furthermore, the coupling of the heavy quark to the plasma particles involves  $g_s^2/4\pi$ multiplied by the Casimir factor $C_F=4/3$. We shall absorb this factor $C_F$ into the coupling $g$,  denoting the product $g_s^2 C_F/4\pi$ by $\alpha=g^2/4\pi$. Thus a coupling constant $\alpha=0.4$ in our notation, corresponds effectively to $\alpha_s=0.3$ in QCD.

The propagator  $\tilde\Delta^{-1}(z,\q)$ introduced above contains all the information about the screening phenomena and the response of the medium to the presence of the heavy quark. It differs by a sign from the longitudinal gluon propagator in the HTL approximation (called $\Delta_L$ in Ref.~\cite{bbr}). It is convenient to subtract from the latter the instantaneous Coulomb interaction which would contribute here only to the self-interaction of the heavy quark. Thus we define
\beq\label{eq:subtDelta}
\Delta(z,\q)=-\left(  \frac{1}{\q^2+\Pi(z,\q)}-\frac{1}{\q^2}  \right).
\eeq
This new object $\Delta(z,\q)$ is proportional to $\chi(z,\q)$, the density-density correlation function of the medium:  $\Delta(z,\q)=(1/q^4)\chi(z,\q)$. 
\begin{figure}[!tp]
\begin{center}
\includegraphics[clip,width=0.7\textwidth]{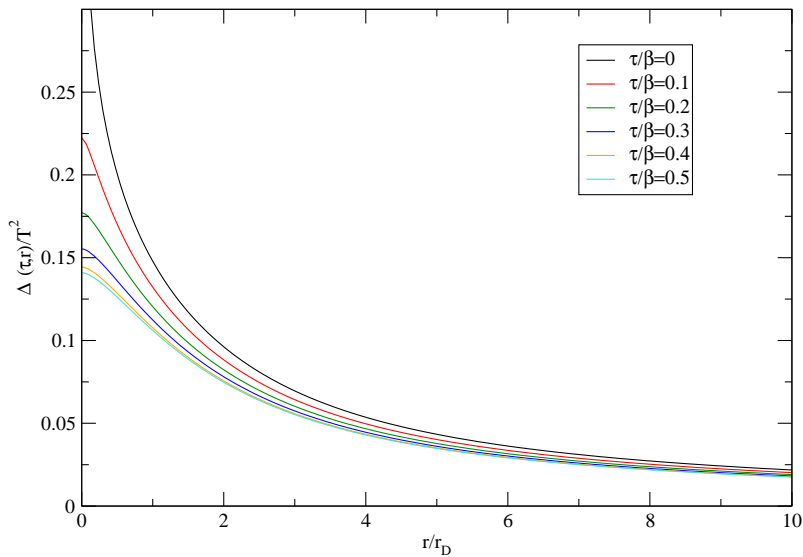}
\caption{The function $\Delta(\tau,\r)$ as a function of $r/r_D=rm_D$ for different values of $\tau/\beta$ (decreasing from bottom to top ). Note that as long as $\tau\ne 0$, $\Delta(\tau, \r=0)$ is finite. However, $\Delta(0, \r)$ diverges logarithmically as $\r\to 0$.}
\label{fig:Delta} 
\end{center}
\end{figure}
One has, in a mixed representation:
\beqa
\Delta(\tau,\q)=
\int_{-\infty}^{+\infty}\frac{dq_0}{2\pi}e^{-q_0\tau}\rho_L(q_0,\q)
[\theta(\tau)+N(q^0)],
\label{eq:realPhi}
\eeqa
where the spectral function $\rho_L(\omega,\q)$ reads \cite{bbr}
\beq\label{eq:HTL_long_spec}
\rho_L(\omega,q)\equiv 2\pi\left\{Z_L(q)\left[\delta(\omega\!-\!\omega_L(q))
-\delta(\omega\!+\!\omega_L(q))\right]
+\theta(q^2\!-\!\omega^2)\beta_L(\omega,q)\right\}. 
\eeq
It displays two types of contributions: A \emph{continuum} term arising from the imaginary part
developed by the logarithm in Eq.~(\ref{polar}) for space-like momenta, and which corresponds physically to scattering processes, 
and a \emph{pole} term, coming from the solution, for time-like momenta, of
\beq\label{eq:dispe}
\q^2+\Pi(\omega_L(\q),\q)=0,
\eeq
which corresponds to an undamped plasma
oscillation. 
Note that the residue $Z_L(q)$ quickly dies out as $q$ grows beyond $m_D$: 
\beq\label{residue}
Z_L(q)\sim \frac{4q}{m_D^2}\exp\left( -\frac{2q^2+m_D^2}{m_D^2}\right), \qquad q\gg m_D.
\eeq
Collective modes exist only for $q\simle m_D$.

The approximation that we are using to describe the medium to which the heavy quark is coupled is motivated by its simplicity, and also by the fact that it encompasses  the important physical phenomena that characterizes weakly coupled plasmas, and that one wants to take into account: polarization and screening effects, collisions with the plasma particles. The latter, however, are not treated properly in the HTL approximation, and this will introduce (small) unphysical features in our results. As a  concrete illustration of the difficulty we are referring to, consider the function $\Delta(\tau,\r)$ that will play a central role in our calculations. This function can be obtained by a Fourier transform of Eq.~(\ref{eq:realPhi}) over spatial momentum, and it is displayed in Fig.~\ref{fig:Delta}. As indicated in the caption of this figure,  $\Delta(0,\r)$ is logarithmically divergent as $\r\to 0$. This divergence is that of  the fluctuation $\langle A_0^2\rangle=\Delta(0,0)$, given by  the integral
\beqa
\langle A_0^2\rangle&=&\int\frac{d^4q}{(2\pi)^4}\rho_L(q^0,q)N(q^0),
\eeqa
and comes form the continuum part of the spectral function (the contribution of the plasmon 
is finite, due to the vanishing of the residue for large wave-numbers; see Eq.~(\ref{residue})). As already mentioned, the continuum part of the spectral function describes scattering processes, involving space-like gluons with energy $\omega$, momentum 
$\q$. In the HTL approximation, the phase space for these  processes is given by $|\omega|\leq q$ (see Eq.~(\ref{eq:HTL_long_spec})), i.e., it grows without bound as $q$ increases, leading eventually to a divergence. An analogous divergence also occurs in the pair correlation function at short distance when this is calculated using the Vlasov equation (which is equivalent to the HTL approximation \cite{pr}). This  is a well known difficulty in plasma physics (see e.g. \cite{Ichimaru}), and it can be cured by a better treatment of the collisions involving large momentum transfer. Indeed, the  HTL approximation is valid only in the regime where $q\ll p$, where $p\sim T$ is a typical loop momentum (i.e., the typical momentum of the colliding plasma particles).  A proper treatment of the collisions with $q\sim p$ would lead to a modified phase space and a finite value for $\langle A_0^2\rangle$.  For instance, in a full one-loop calculation, the phase space is given by $-q<\omega<q$ for $q\simle p$, but $q-2p<\omega<q$ for $q>p$. A possible way to improve the calculation would be to introduce a cut-off to separate soft and hard contributions, and apply in each sector appropriate approximations. We shall not do so here, because $\Delta(t,\r)$ enters the calculation of the heavy quark  correlator only through an integral so that its logarithmic singularity is tamed, and its physical consequences mild. We note however that  the divergence of $\langle A_0^2\rangle$ modifies the small $\tau$ behavior of the heavy quark propagator, and in particular it invalidates the Taylor expansion dicussed at the end of Sect.~\ref{sec:generalities}, beyond the linear order. 
%

\subsection{Path integral representation}
\label{sec:path_integral}

We are now in position to write the propagator of the heavy quark in the form of a path integral. Gathering the results of the first two sections, we can write
\beqa\label{pathintegral0}
 G^>(-i\tau,\r)=\int {\cal D} A_0^E\,\exp\left[  -\frac{1}{2}\int_{x,x'} A_0^E(x)\tilde\Delta^{-1}(x,x')A_0^E(x')\right] 
\qquad\qquad\nonumber\\
\times \int_{0}^{\r} {\cal D}\x\,\exp\left[   -\int_0^\tau d\tau' \left( \frac{1}{2}M\dot\x^2 +ie A^E_0(\tau,\x)  \right)\right],
\eeqa
This path integral summarizes the model that we use. The dynamics of the heavy quark in a hot plasma is that of a non relativistic quantum particle moving in a fluctuating field $A_0$, and this is treated exactly by the path integral. The approximations only enter the description of the fluctuations of the field $A_0$ which we assume to be Gaussian and, as we have just discussed, dominated by long wavelengths and low frequencies. Thus, any improvement of the description of the medium will affect only the first part of the functional integral (\ref{pathintegral0}), that is the weight of the integration over the field $A_0$, but it will  leave intact  the second part describing the motion of the heavy quark in the fluctuating field. This is an important feature of the present description. 

As we mentioned earlier, it is convenient to subtract from the correlator $\tilde\Delta$ the contribution of the Coulomb interaction. This  is most easily done after having performed the Gaussian integral over $A_0^E$, whence we can just replace $\tilde \Delta$ by $-\Delta$. One gets
\beq\label{eq:1part_pathQED}
G^>(-i\tau,\r)
\!=\!\int_{0}^{\r}{\cal D}
\z\,\, {\rm e}^{-S[\z,\tau]},
\eeq
where $S[\z,\tau]=S_0[\z,\tau]-\bar F[\z,\tau]$, with
\beq
S_0[\z,\tau]=\int_{0}^{\tau}d\tau'\frac{1}{2}M\dot{\z}^2, 
\eeq
and
\beq\label{barF}
\bar F[\z,\tau]=\frac{{g}^2}{2}\int_0^\tau d\tau' \int_0^\tau d\tau''
\Delta(\tau'-\tau'',\z(\tau')-\z(\tau''))
.
\eeq
The real time version of this  path integral is obtained by replacing $\tau$ by $it$, and substituting $-S[\z,\tau]$ in Eq.~\ref{eq:1part_pathQED} by $iS[\z,t]=i(S_0+F)$ with 
\beq\label{Fdezett}
F[\z,t]=\frac{{g}^2}{2}\int_0^t dt' \int_0^t dt''
D(t'-t'',\z(t')-\z(t''))
,
\eeq
and we have used \cite{bbr}
\beq\label{DDelta}
\Delta(\tau=it,\r)\equiv -iD(t,\r).
\eeq

The correlator  $G^>(t,\r)$, when expressed in terms of the dimensionless variables $tT$ and $\r T$ is of the form 
 $G^>(t,\r)=T^3 f(M/T, m_D/T, tT, \r T)$, with $f$ a dimensionless function. When $m_D\to 0$ this reduces to the free propagator.  Note that at fixed value of the coupling, $m_D/T$ is fixed, and $G^>(t,\r)$, when $ t$ and $r$ are expressed in units of the inverse temperature, depends only on the ratio $M/T$.  We shall refer to this scaling property of the correlator repeatedly. 
 
 The parameter $M/T$ controls the ``diffusion'', described  by the correlator (\ref{freesolution}) in imaginary time: the smaller $M/T$, the more the heavy quark will move away form the origin in a given time.  Note that this diffusion inhibits the effects of the interaction: because $\Delta(\tau,\z)<\Delta(\tau,0)$ (see Fig.~\ref{fig:Delta}), the interaction favors paths for which $\z$ remains small (their weight in Eq.~(\ref{eq:1part_pathQED}) is largest). 
 
 One may also understand the effect of the interaction in the following way. The heavy quark produces a polarization cloud of light particles around itself. This induced charge screens that of the heavy quark over a distance scale of order $m_D^{-1}$.  When the heavy quark moves, its polarization cloud tries to adjust and follow its motion, but this takes time (of order $m_D^{-1}$). The heavy quarks sees then a restoring force produced by the lagging polarization cloud, which limits its excursion.  
 
 In the limit $M/T\to \infty$, studied in detail in the next subsection, the heavy quark is frozen at it initial location: there is then no diffusion, and the effect of interactions is maximal.

\subsection{The infinite mass limit}

When $M\to\infty$, the path integral can be calculated exactly. This is because, in this limit, the motion of the heavy quark is frozen and $F$ becomes independent of the coordinates. Thus, in the infinite mass limit, the heavy quark correlator takes the form
\beq\label{GFlargeM}
G^>(t,\r)
=\delta(\r) \,{\rm e}^{-iMt}  \,{\rm e}^{iF(t)}\eeq
where the function $F(t)$ is the functional (\ref{Fdezett}) restricted to $\z=0$: 
\beq
F(t)=\frac{{g}^2}{2}\int_0^t dt' \int_0^t dt''
D(t'-t'',0)
.\label{eq:1part_pathQEDMinf}
\eeq
The factor $\exp \left( iF(t)\right) $ mulitplying in Eq.~(\ref{GFlargeM}) the infinite mass limit of the free correlator (\ref{freesolution}), summarizes the effect of the interactions.
One can express $F(t)$ in terms of the Fourier transform $D(\omega,\q)$ of the (time-ordered) gluon  propagator  \cite{bbr}:
\beq\label{eq:realtFourier}
D(\omega,\q)=\int\frac{dq^0}{2\pi}\frac{\rho_L(q^0,\q)}{q^0-(\omega+i\eta)}+i\rho_L(\omega,\q)N(\omega).
\eeq
One gets
\beq
F(t)=g^2\int\frac{d\omega}{2\pi}\,\frac{1-\cos(\omega t)}{\omega^2}\,\int\frac{d^3\q}{(2\pi)^3}\,D(\omega,\q).
\eeq
It follows that at short times
\beq
F(t)\simeq \frac{g^2}{2} t^2 \int\frac{d\omega}{2\pi}\int\frac{d^3\q}{(2\pi)^3}D(\omega,\q)=\frac{g^2}{2} t^2 D(t=0,\r={0}).
\eeq
For large time we use
\beq
\lim_{t\to\infty}\frac{1-\cos(\omega t)}{\omega^2}=\pi t \delta(\omega).
\label{eq:limcos}
\eeq
to obtain
\beq\label{Flargetime}
F(t)\simeq \frac{g^2}{2} t D(\omega=0,\r={0})\equiv -t V_{opt}.
\eeq
An alternative way to obtain this result is to  start directly from Eq.~(\ref{eq:1part_pathQEDMinf}) and  to change variables $t'-t''\to u, (t'+t'')/2\to T$, and to observe that at large time $t$, one may integrate freely over $u$, thereby filtering out the zero frequency part of $D(\omega,\q)$. This yields again Eq.~(\ref{Flargetime}).
Thus, the large time ($t\gg m_D^{-1}$) behavior of the system is determined by the static ($\omega=0$) response of the medium. Since, at large times, $F$ is linear in time, Eq.~(\ref{eqmG>}) for $G^>(t,\r=0)$ is a closed equation that takes the form of a Schr\"odinger equation \cite{bbr}, with an ``optical potential'' $V_{opt}$ given by 
\beqa\label{eq:vopt}
V_{\rm opt}&\equiv&- \frac{{g}^2}{2}\int\frac{d\q}{(2\pi)^3} D(\omega\!=\!0,\q)
\nonumber\\
{}&=&\frac{{g}^2}{2}\int\frac{d\q}{(2\pi)^3}\Big[\frac{1}{\q^2+m_D^2}-\frac{1}{\q^2} \!-i\frac{\pi m_D^2 T} {|\q|(\q^2+m_D^2)^2}\Big]\nonumber\\
{}&=&-\frac{\alpha}{2} m_D
-i\frac{\alpha T}{2},
\eeqa
where we have used the susceptibility sum rule
\beq
\int_{-\infty}^{\infty}\frac{dq_0}{2\pi}\,\frac{\rho_L(q_0,\q)}{q_0}=\frac{m_D^2}{\q^2(\q^2+m_D^2)}
\eeq
in order to perform the $q_0$ integral needed to calculate $D(\omega=0,\q)$. 
Thus, the optical potential involves a real correction to the mass of the heavy quark, and an imaginary part  that takes into account the coupling of the heavy quark to the complex configurations of the medium. Alternatively, one may view this imaginary part as reflecting  the collisions of the heavy quark with the particles of the medium. As we shall see in the next section, $V_{\rm opt}$ can be identified with the one-loop on-shell self energy in the infinite mass limit. 

 This imaginary part does not appear in the Euclidean correlator calculated at $\tau=\beta$, which exhibits only the mass shift \cite{bbr}:
\beq\label{GbetaDF}
-T\ln G^>(t=-i\beta,\p)= M-\frac{\alpha}{2} m_D=M-\frac{g^2}{2}\Delta(i\omega_n=0,\r=0).
\eeq
Note that in the present, infinite mass, limit, $G^>(t=-i\beta,\p)$ in fact does not depend on $\p$.
More generally,  the Euclidean correlator has the form (\ref{GFlargeM}) with $iF(t)$ replaced by $\bar F(\tau)$, with (see Eq.~(\ref{barF})):
\beq\label{eq:Ftau}
\bar F(\tau)=\frac{{g}^2}{2}\int_0^\tau d\tau' \int_0^\tau d\tau'' \Delta(\tau'-\tau'',\0).\eeq
Since the dominant effect of the interactions can be characterized by the free energy shift (\ref{GbetaDF}), it is convenient to separate the corresponding linear growth of $\bar F(\tau)$, and write
\beq\label{eq:1e2}
\bar F(\tau)= \bar F_1(\tau)+\bar F_2(\tau),
\eeq
with
\beq
\bar F_1(\tau)=\frac{g^2}{2} \tau\int\frac{d^3\q}{(2\pi)^3} \int\frac{dq_0}{2\pi}\frac{\rho_L(q_0,\q)}{q_0}=\frac{\tau}{\beta} \bar F(\beta),
\eeq
and
\beq\label{eq:m_inf_f}
\bar F_2(\tau)=\frac{g^2}{2}\int\frac{d^3\q}{(2\pi)^3}\int\frac{dq_0}{2\pi}\frac{\rho_L(q_0,\q)}{q_0^2}\frac{\cosh(q_0(\tau-\beta/2))-\cosh(\beta q_0/2)}{\sinh(\beta q_0/2)}.
\eeq
The function $\bar F_2(\tau)$ vanishes at $\tau=0$ and $\tau=\beta$. It is symmetric around $\tau=\beta/2$, a property that follows immediately from the fact that $\Delta(\tau,\q)$ depends only on $|\tau|$, and is periodic, $\Delta(\beta,\q)=\Delta(0,\q)$. Note also that the slope of $\bar F_2(\tau)$ at the origin is equal and opposite to that of $\bar F_1(\tau)$, since that of $\bar F(\tau)$ vanishes.

Many of (but not all) the features of the present $M\!\to\!\infty$ limit are shared by the toy model
presented in Appendix~\ref{sec:toy-model}, where one can find a more extended discussion of some of the points addressed in this subsection.
\section{One-loop calculation}\label{sec:physical_processes}

In this section, we present the results of the one-loop calculation of the heavy quark correlator. This provides insight into the dynamics of the heavy quark when the interaction is weak enough for perturbation theory to be applicable.
All the numerical results to be presented are obtained with the value $\alpha=0.4$ of the coupling constant, which appears to be a moderate value for which the one-loop approximation remains reasonably accurate. This calculation, together with the exact large $M$ limit that we have just discussed, will serve as a reference when discussing  the results of the Monte Carlo evaluation of the heavy-quark correlator  in the next section.

\begin{figure}[!tp]
\begin{center}
\includegraphics[clip,height=4cm]{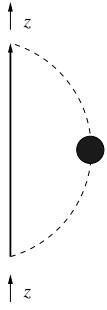}
\caption{The one-loop self-energy diagram for the heavy quark. The blob on the interaction line reminds that the latter represents a resummed HTL propagator of a longitudinal gluon.}\label{fig:diagram}
\end{center}
\end{figure}

The one-loop calculation is easier in momentum space than in coordinate space. To proceed we consider the analytic propagator 
\beq\label{Gdez}
G(z,\p)=\frac{-1}{z-E_p-\Sigma(z,\p)},
\eeq
where $E_p\!=\! M\!+\!\p^2/2M$, and the one-loop self-energy $\Sigma(z,\p)$ is given by the diagram displayed in Fig.~\ref{fig:diagram}.
The retarded propagator is obtained as usual by setting $z\!=\!\omega\!+\!i\eta$, with $\omega$ real. The imaginary part of the retarded propagator yields the  heavy-quark spectral function
\beq\label{eq:HQspectral_def}
\sigma(\omega,\p)\equiv 2{\rm Im}\,G^R(\omega,\p)=\frac{\Gamma(\omega,\p)}{[\omega-E_p-{\rm Re}\,\Sigma(\omega,\p)]^2+\Gamma^2(\omega,\p)/4}, 
\eeq
where $\Gamma(\omega,\p)\!\equiv\!-2{\rm Im}\,\Sigma^R(\omega,\p)\!=\!-2{\rm Im}\,\Sigma(z\!=\!\omega\!+\!i\eta,\p)$. Eventually the Euclidean correlator will be calculated using Eq.~(\ref{euclcorrelspec}). Since  we  shall consider only the case $\p=0$, we shall  use in most of this section the simplified notation $\Sigma(z)$ for  $\Sigma(z,\p=0)$, and similarly for other related functions. The relations (\ref{Gdez}) and (\ref{eq:HQspectral_def}) are general, but in the rest of this section, $G$ and $\Sigma$ will refer to one-loop quantities (unless stated otherwise). 

\subsection{The one-loop self-energy}

A standard calculation, implementing  approximations that are valid when $T/M\ll 1$, yields the  analytic one-loop self-energy 
 \beq\label{eq:test}
\Sigma(z)=g^2\int\frac{d\k}{(2\pi)^3}\int_{-\infty}^{+\infty}\frac{dk^0}{2\pi}\rho_L(k^0,k)\frac{1+N(k^0)}{z-E_k-k^0}.
\eeq
Expressing momenta and energies in units of $T$, one sees that $\Sigma(z)$ is a function of the form $\Sigma(z)=T f(z/T,M/T,m_D/T)$. At fixed value of the coupling constant, $m_D/T$ is a constant, so that,  the only relevant parameter is the ratio $T/M$, as we have already mentioned.

\begin{figure}[!tp]
\begin{center}
\includegraphics[clip,width=0.7\textwidth]{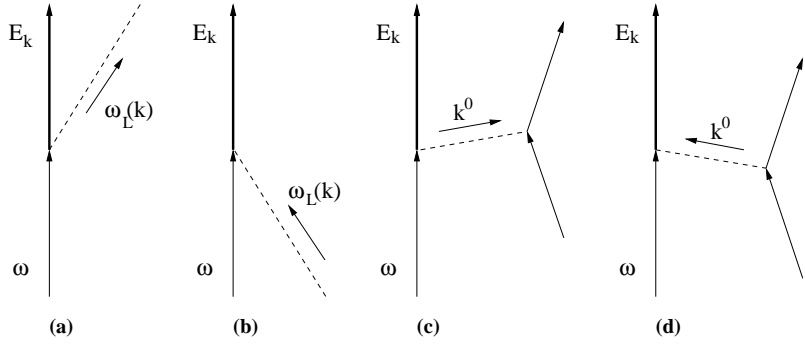}
\caption{The different processes contributing to the imaginary-part of the heavy-quark self-energy: (a)-(b) emission-absorption of a plasmon and (c)-(d) collisions with the plasma particles, mediated by one-gluon exchange.}
\label{fig:diagrams} 
\end{center}
\end{figure}

By using the explicit expression for the gluon spectral function $\rho_L(k^0,k)$ given in Eq.~(\ref{eq:HTL_long_spec}),  one can re-write Eq.~(\ref{eq:test}) as 
\begin{multline}\label{Sigmadez}
\Sigma(z)=g^2\int\frac{d\k}{(2\pi)^3}\left\{Z_L(k)\left[\frac{1+N(\omega_L(k))}{z-E_k-\omega_L(k)}+
\frac{N(\omega_L(k))}{z-E_k+\omega_L(k)}\right]+\right.\\
\left.+\int_{0}^{k}\frac{dk^0}{2\pi}\,2\pi\,\beta_L(k^0,k)\left[\frac{1+N(k^0)}{z-E_k-k^0}+
\frac{N(k^0)}{z-E_k+k^0}\right]\right\}.
\end{multline}
This expression exhibits two types of contributions that are  are illustrated in  Fig.~\ref{fig:diagrams}: a pole contribution whose energy denominators are associated with processes of emission or absorption of collective plasmons by the heavy quark, and a continuum contribution coming from the continuum part of the gluon spectral density; the latter contribution represents the effect of collisions between the heavy quark and the particles of the medium, mediated by space-like gluons. 
It is convenient to  evaluate separately these two contributions. Accordingly, we set $\Gamma(\omega)=\Gamma^{\rm pole}(\omega)+\Gamma^{\rm cont}(\omega)$. 

\begin{figure}[!tp]
\begin{center}
\includegraphics[clip,width=0.60\textwidth]{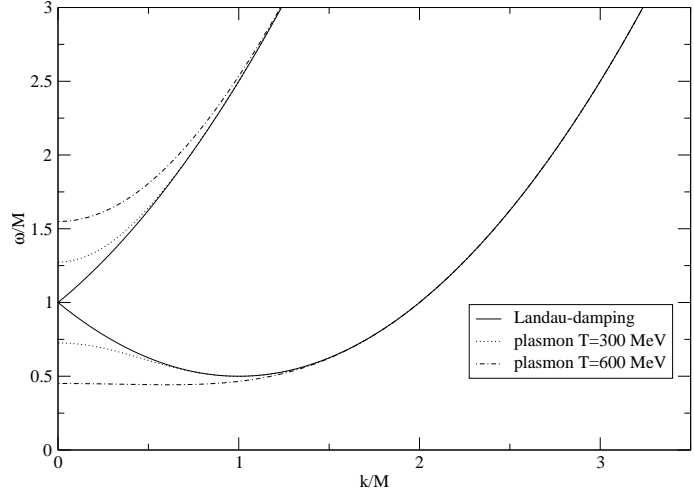}
\caption{The dashed curves represent, for two different temperatures, the functions $E_k\!\pm\!\omega_L(k)$, with $E_k\!=\!M\!+\!\frac{\k^2}{2M}$, and $\omega_L(k)$ the plasma dispersion relation. For $k\!=\!0$, $\omega\!=\!M\!\pm\! \omega_{\rm pl}$, with $\omega_{\rm pl}\!=\!m_D/\sqrt{3}$ the plasma frequency, proportional to the temperature. Here, $M\!=\!1.5$ GeV and $\omega_{\rm pl}\!=\!412$ MeV for $T\!=\!300$ MeV. The two full lines delineate the support of the continuum part of the gluon spectral function, that is the region $-k\!\le\! \omega\!-\!M\!-\!\frac{\k^2}{2M}\!\le\! k$. The largest temperature $T\!=\!600$ MeV corresponds to a plasma frequency $\omega_{\rm pl}\!=\!824$ MeV, very close to the ``critical value'' $3/5M$ discussed in the text.}
\label{fig:response_tot} 
\end{center}
\end{figure}

\begin{figure}[!tp]
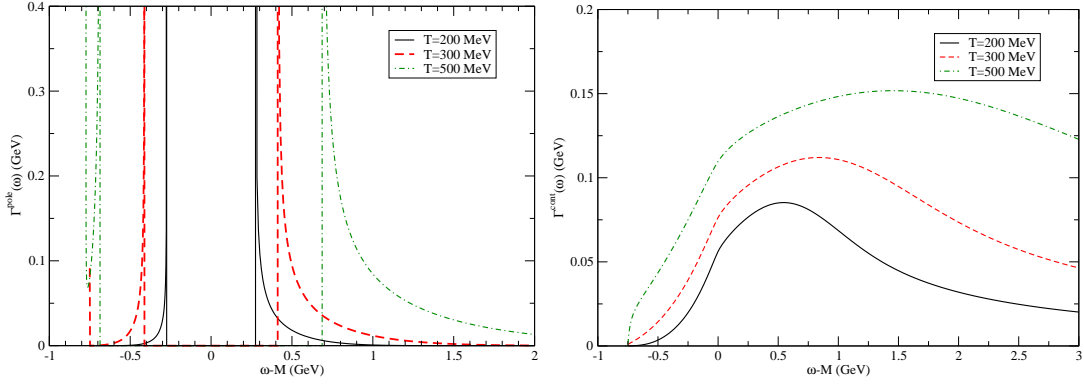

\begin{center}
\includegraphics[clip,width=0.47\textwidth]{gammapole_charm.eps}
\includegraphics[clip,width=0.47\textwidth]{gammacont_charm.eps}
\caption{The pole (left) and continuum (right) contribution to $\Gamma(\omega)$ for a quark mass $M\!=\!1.5$ GeV, and various temperatures. The plasma frequency is proportional to the temperature and has values $\omega_{\rm pl}\!=\!275$, 412 and 687 MeV for the temperatures $T\!=\!200$, 300 and 500 MeV, respectively. Notice, in the pole contribution,  the Van-Hove singularities and the gap, for $M-\omega_{\rm pl}\!<\!\omega\!<\!M+\omega_{\rm pl}$, that increases with temperature. The continuum contribution grows linearly with temperature, and  the threshold at $\omega=M/2$ is clearly visible.}
\label{fig:gamma_polecont} 
\end{center}
\end{figure}

For the pole contribution one gets:
\begin{multline}\label{eq:pole_num}
\Gamma^{\rm pole}(\omega)=\frac{g^2}{\pi}\left\{\frac{k_1^2}{|E_{k_1}'+\omega_L'(k_1)|}Z_L(k_1)\left[1+N(\omega_L(k_1))\right]+\right.\\
+\left.\sum_{k_2}\frac{k_2^2}{|E_{k_2}'-\omega_L'(k_2)|}Z_L(k_2)N(\omega_L(k_2))
\right\},
\end{multline}
where $k_1$ and $k_2$ are implicit functions of $\omega$ given by 
\beq\label{eq:polesol}
\omega=E_{k_1}+\omega_L(k_1),\quad\quad\omega=E_{k_2}-\omega_L(k_2),
\eeq
and the primes in the denominators of Eq.~(\ref{eq:pole_num}) denote derivatives with respect to $k$.
Here $\omega_L(k)$ is the plasmon dispersion relation (see Eq.~(\ref{eq:dispe})),
whose behavior for small momenta
reads:
\beq\label{eq:plasm_small}
\omega_L^2(k)\underset{k\ll m_D}{\sim}\omega_{\rm pl}^2+\frac{3}{5}k^2\quad\Rightarrow\quad
\omega_L(k)\underset{k\ll m_D}{\sim}\omega_{\rm pl}+\frac{3}{10}\frac{k^2}{\omega_{\rm pl}} .
\eeq
The solutions of  Eqs.~(\ref{eq:polesol}) can be read out from Fig.~\ref{fig:response_tot}
where the two curves $E_k\!\pm\! \omega_L(k)$ are plotted as a function of $k$.
The first equation (\ref{eq:polesol}), $\omega=E_{k}+\omega_L(k)$, has a single solution starting from
the plasmon-emission threshold $\omega\!=\!M\!+\omega_{\rm pl}$.
The number of solutions of the second equation depends
on the ratio $\omega_{\rm pl}/M$. From Eqs.~(\ref{eq:polesol}) and (\ref{eq:plasm_small}) one sees
that for $\omega_{\rm pl}\!>\!\frac{3}{5}M$ the dispersion relation $\omega=E_{k}-\omega_L(k)$ starts with positive
curvature and it  contributes to $\Gamma^{\rm pole}$ with a single solution starting from
$\omega\!>\!M\!-\omega_{\rm pl}$. In the case of interest, $T/M\!\ll\! 1$,
we have $\omega_{\rm pl}\!<\!\frac{3}{5}M$, and there are two solutions
for $M/2\!\simle\!\omega\!<\!M\!-\!\omega_{\rm pl}$ and only one for $\omega\!>\!M\!-\omega_{\rm pl}$.
Note however that  any contribution corresponding to $k\!\sim\! M$ is
damped by the plasmon-residue. As a result, there is  effectively no pole term contribution
for $M\!-\!\omega_{\rm pl}\!<\!\omega\!<\!M\!+\!\omega_{\rm pl}$, as it can be seen from 
the left panel of Fig.~\ref{fig:gamma_polecont}.
From Fig.~\ref{fig:response_tot} one realizes also that there are values of $\omega$
for which
the dispersion relations $\omega\!=\!E_k\!\pm\!\omega_L(k)$
display stationary points which, because of the denominators in Eq.~(\ref{eq:pole_num}), lead to singularities in $\Gamma^{\rm pole}(\omega)$ (\emph{Van-Hove singularities}),  clearly visible in the left panel of Fig.~\ref{fig:gamma_polecont}.
Notice that, as the ratio $T/M$ gets larger, the one occurring at $\omega\!\sim\! M/2$
acquires more importance, getting less suppression from the plasmon residue.  

The continuum contribution involves the spectral density $\beta_L(k_0, \k)$ which has support for $|k_0|\le k$. It follows then from Eq.~(\ref{Sigmadez}) that the continuum contribution to the imaginary part of $\Sigma$ comes from  values of $k,\omega$ such that $-k\le \omega-M-\frac{\k^2}{2M}\le k$. The boundaries of this domain are displayed in Fig.~\ref{fig:response_tot}, and $\Gamma^{\rm cont}(\omega)$ is given by
\begin{multline}
\Gamma^{\rm cont}(\omega)=\frac{g^2}{\pi}\!
\left\{
\theta(\omega\!-\!M)\int_{-M+\overline{M}(\omega)}^{M+\overline{M}(\omega)}k^2\,dk\,
\beta_L(\omega\!-\!E_k,k)\left[1\!+\!N(\omega\!-\!E_k)\right]
+\right.\\
\left.
+\,\theta(\omega\!-\!M/2)\,\theta(M\!-\!\omega)\!\int_{M-\overline{M}(\omega)}^{M+\overline{M}(\omega)}k^2\,dk\,
\beta_L(\omega\!-\!E_k,k)\left[1\!+\!N(\omega\!-\!E_k)\right]
\right\},
\end{multline}
where $\overline{M}(\omega)\!\equiv\!\sqrt{M^2\!+\!2M(\omega\!-\!M)}$. Note in particular the lower threshold at $\omega=M/2$, corresponding to the minimum at $k=M$ of the lower boundary of the support displayed in Fig.~\ref{fig:response_tot}. This is clearly visible in the plot of  $\Gamma^{\rm cont}(\omega)$ in the right hand panel of Fig.~\ref{fig:gamma_polecont}.

\begin{figure}[!tp]
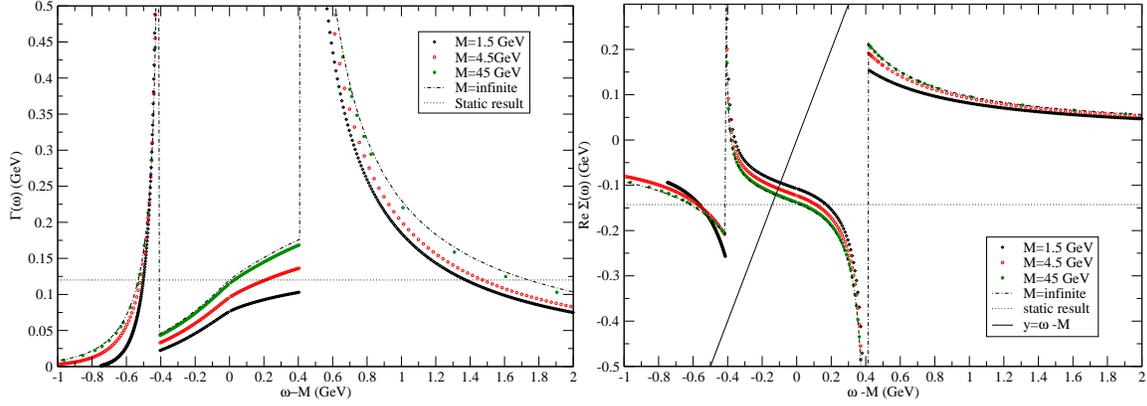

\begin{center}
\includegraphics[clip,width=0.5\textwidth]{imsigma_complete.eps}\includegraphics[clip,width=0.5\textwidth]{resigma_complete.eps}
\caption{Imaginary part (left)  and real part (right) of the self-energy $\Sigma$. The horizontal lines, labelled ``static limit'', indicate the values of $\Gamma(M\!\to\!\infty)$ and ${\rm Re}\Sigma(M\!\to\!\infty)$. With the parameters $\alpha\!=\!0.4$ and $T\!=\!300$ Mev, these are respectively 120 MeV and -143 MeV. Within the gap $\pm \omega_{pl}$, $\Gamma $ is an increasing function of $M$, while ${\rm Re}\Sigma$ is a decreasing function of $M$. Both functions nearly reach the infinite mass limit when $M=45$ GeV.}
\label{fig:ImRe} 
\end{center}
\end{figure}
A similar analysis can be done for the real part of the self-energy. This will not be detailed here. We just present in Fig.~\ref{fig:ImRe} the result of the full calculation of the imaginary part (left panel) and the real part (right panel)  of $\Sigma$, for different values of the heavy quark mass, including the limiting case of infinite mass. Let us recall that in the latter case, we have analytic results  \cite{bbr} for the on-shell values (corresponding to $\omega=M$). From Eq.~(\ref{Sigmadez}) one sees that only the continuum part contributes to the on-shell imaginary part
\beq\label{eq:im_inf}
{\rm Im}\Sigma^R(M\!\to\!\infty)\!=\!-\frac{g^2}{2}\lim_{k^0\to 0}\int\frac{d\k}{(2\pi)^3}N(k^0)\rho_L(k^0,k)\!=\!-\,\frac{\alpha T}{2}\quad\Rightarrow\quad \Gamma(M\!\to\!\infty)\!=\!\alpha T ,
\eeq
while the real part receives contribution from both parts of $\Sigma(z)$:
\beq
{\rm Re}\Sigma(\!M\!\to\!\infty)=-\frac{g^2}{2}\int\frac{d\k}{(2\pi)^3}\int_{-\infty}^{+\infty}\frac{dk^0}{2\pi}\frac{\rho_L(k^0,k)}{k^0}=-\,\frac{\alpha m_D}{2}.
\eeq
These values, which coincide with the values obtained for the exact ``optical potential'' in Eq.~(\ref{eq:vopt}),  are indicated by the horizontal lines (labelled as ``static result'') in Fig.~\ref{fig:ImRe}, while the curves representing the full expressions of ${\rm Re}\Sigma(\omega)$ and $\Gamma(\omega)$ in the infinite mass limit are labelled as ``M=infinite''. One sees from this figure that the infinite mass limit is nearly attained for $M=45$ GeV, and that finite mass effects do not change the qualitative behavior of the self-energy. To get a quantitative measure of these finite mass effects, we determine the shift $\delta M=M'-M$ of the heavy quark mass as given by the solution of the equation
\beq\label{DysonM}
M'-M={\rm Re}\Sigma(M').
\eeq
This can be obtained graphically, as the intersection of the line $\omega-M$ with $\Sigma(\omega)$ in  Fig.~\ref{fig:ImRe}. Values of the mass shift $\delta M$ thus obtained are reported in Table~\ref{massshift}.

\begin{table}[htdp]
\begin{center}
\begin{tabular}{|c|c|c|c|c|c|c|}
\hline
$T/M$&0&0.0067&0.067&0.133&0.200& 0.333\\
\hline
\hline
$\delta M/T$&  -0.407 &-0.4 &-0.357 &-0.335&-0317&-0.288  \\
\hline
$\Delta F_Q/T$ &-0.416 &-0.409 &-0.362 &-0.336& -0.318&-0.274 \\
\hline
${\rm Re}\Sigma(M)/T$ &-0.476 &-0.457 &-0.41 &-0.38&-0.357&-0.326\\
\hline
\end{tabular}
\caption{The mass shift $\delta M$ obtained from the solution of Eq.~(\ref{DysonM}), the one-loop free energy shift $\Delta F_Q$, and the real part of the on-shell self-energy ${\rm Re}\,\Sigma(M)$ (which equals the exact energy shift in the infinite mass limit), as a function of $T/M$.}
\end{center}
\label{massshift}
\end{table}
One sees from this table 
that the larger the ratio $T/M$, the smaller the mass-shift. This is in line with what one expects from the effects of diffusion that increase as $T/M$ increases, and inhibits the effect of the interaction. Note also that the mass shift obtained as the solution of Eq.~(\ref{DysonM}) is numerically very close to the free energy shift calculated from the Euclidean correlator $G(-i\beta)$. It is smaller (in absolute value) than ${\rm Re}\,\Sigma(\omega=M)$, as can be also directly seen in Fig.~\ref{fig:ImRe}.  
%
\subsection{One-loop spectral function and Euclidean correlator}\label{sec:goneloop}
\begin{figure}[!tp]
\begin{center}
\includegraphics[clip,width=0.7\textwidth]{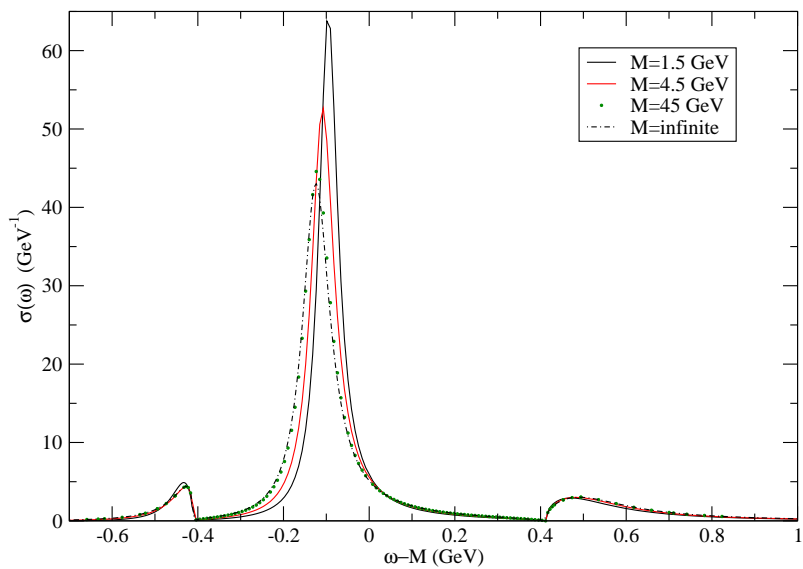}
\caption{One-loop spectral function as a function of $\omega\!-\!M$, for various values of the heavy quark mass, and a fixed  tempearture $T\!=\!300$ MeV. The curve corresponding to $M\!=\!45$ GeV is hardly distinguishable from that representing the one-loop infinite mass limit. The smaller the mass $M$, the smaller the shift of the main peak.}
\label{fig:spectral_oneloop} 
\end{center}
\end{figure}
The spectral density can be readily calculated from  the real and imaginary parts of the self-energy (see Eq.~(\ref{eq:HQspectral_def})). It is displayed in Fig.~\ref{fig:spectral_oneloop}. The dominant feature is the existence of a main peak, approximately located at the value of $\omega=M'$, with $M'$ given by   Eq.~(\ref{DysonM}), as can be expected on general grounds from Eq.~(\ref{eq:HQspectral_def}). In addition to the main peak, two secondary bumps appear in the spectrum at values of the energy $\omega\approx M\pm\omega_{\rm pl}$, and come from the energy dependence of the imaginary part of $\Sigma$ discussed in the previous subsection. 
The spectral density satisfies the sum rules (\ref{sumrules012}): it is normalized to $1$, and its first moment remains  equal to $M$.   Note that the infinite mass limit gives an accurate picture, only mildly modified by finite mass corrections, down to values of the mass of the order of 1.5 GeV.
In particular finite $T/M$ effects seem to be important mainly for the shift
and the broadening of the main peak,
affecting on the other hand very mildly the secondary bumps.
\begin{figure}[!tp]
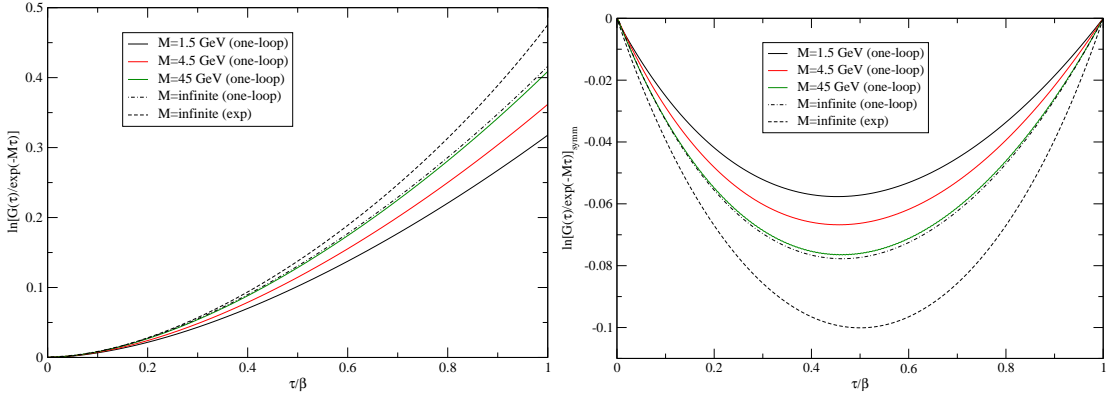

\begin{center}
\includegraphics[clip,width=0.48\textwidth]{gtau_Mscan_arg_complete.eps}
\includegraphics[clip,width=0.48\textwidth]{gtau_symm_Mscan_complete.eps}
\caption{The quantity $\bar F^{(1l)}(\tau)$ (see Eq.~\ref{eq:F1loop})) for various values of the mass and fixed $T\!=\!300$ MeV. For the $M\!=\!\infty$ case we also plot the function $\bar F(\tau)$ (the curve labelled ``exp''), so as to get a measure of the accuracy of the one-loop approximation. The value of the free energy shift $\Delta F$ can be read on the left panel as the value of $\bar F^{(1l)}(\beta)$ and is reported in Table~\ref{massshift}. In the right panel we plot the same quantities after subtracting the linear $\tau$-dependence driven by the free-energy, that is the function $\bar F_2^{(1l)}(\tau)$ in Eq.~(\ref{eq:F1loop12}).}
\label{fig:correlator_one_loop2} 
\end{center}
\end{figure}

By using the relation (\ref{euclcorrelspec}), 
one obtains from the one-loop spectral function the corresponding
Euclidean correlator.
This is displayed in Fig.~\ref{fig:correlator_one_loop2}, for different
values of the heavy quark mass. What is plotted in Fig.~\ref{fig:correlator_one_loop2} is actually the function 
\beq\label{eq:F1loop}
\bar F^{(1l)}(\tau)=\ln\frac{G^{>}(-i\tau)}{G_0(-i\tau)}.
\eeq
As it can be seen, 
all the curves, 
start with zero slope at $\tau=0$. This is related to the general feature that the interactions do not introduce any corrections linear in $\tau$ at small $\tau$, which in turn may be linked to the first two sum rules (\ref{sumrules012})) which are satisfied in the one-loop approximation. This represents actually an important consistency check of the
numerical calculation, given the indirect way by which the Euclidean correlator was obtained.
The value of the Euclidean
correlator at $\tau\!=\!\beta$ 
measures the free-energy shift $\Delta F$ caused by the addition of the heavy quark, and can be read off Fig.~\ref{fig:correlator_one_loop2}.
As already obtained in the case of $\delta M$, one finds a smaller shift as
the ratio $T/M$ gets larger (see Table~\ref{massshift}). 

A different way to plot the Euclidean correlator is offered in the right panel
of Fig.~\ref{fig:correlator_one_loop2}.
There we have separated the linear $\tau$-dependence driven by the free-energy shift, writing (see Eq.~(\ref{eq:1e2}))
\beq\label{eq:F1loop12}
\bar F^{(1l)}(\tau)=\bar F_1^{(1l)}(\tau)+\bar F_2^{(1l)}(\tau),\qquad \bar F_1^{(1l)}(\tau)=\frac{\tau}{\beta}\bar F^{(1l)}(\beta).
\eeq
The difference of behavior that is observed is quite similar to that
obtained in the toy model presented in Appendix~\ref{sec:toy-model}.
Note in particular that the symmetry around $\beta/2$ that is present in the exact $M\!=\!\infty$ limit
is lost in the one-loop approximation (also in the infinite mass limit of the latter).

\begin{figure}[!tp]
\begin{center}
\includegraphics[clip,width=0.7\textwidth]{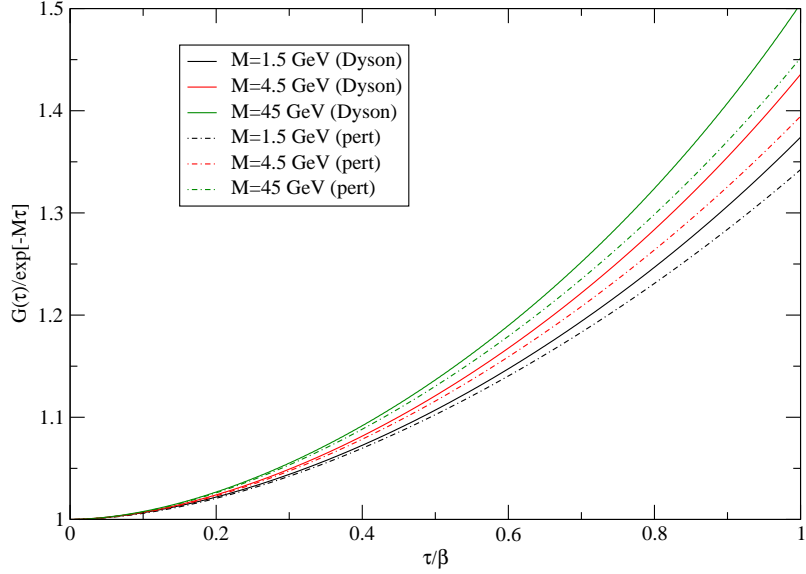}
\caption{A comparison between the first-order perturbative expansion
of the Euclidean correlator $G_{\rm pert}(\tau)\!\equiv\!G_0(\tau)\!+\!G_1(\tau)$
(dashed lines)
and the full one-loop correlator $G(\tau)$ obtained from the resummation of the Dyson series (continuous lines).}
\label{fig:g_smalltau} 
\end{center}
\end{figure}

Finally it is of interest to study the accuracy of the weak-coupling expansion at short time. 
To that aim, we expand the propagator to order $\alpha$
\beq
G(\tau)=G_0(\tau)+G_1(\tau)+\dots,
\eeq
with $ G_0(\tau)\equiv e^{-M\tau}$, and 
 $G_1(\tau)$ is given by the one-loop self-energy:
\beq\label{eq:G10}
G_1(\tau)=g^2{\rm e}^{-M\tau} \!\!\int_0^\tau \!\!\!d\tau'\!\!\int_0^{\tau'} \!\!\!d\tau''\int\frac{d^3\k}{(2\pi)^3}\Delta(\tau'-\tau'',\k)\,{\rm e}^{-(\k^2/2M)(\tau'-\tau'')}.
\eeq
In order to calculate the time intergral, one may express the gluon propagator in terms of its spectral density. One gets then:
\begin{multline}\label{eq:G1}
G_1(\tau)/e^{-M\tau}=g^2\int\frac{d\k}{(2\pi)^3}\int\frac{dk^0}{2\pi}\frac{\rho_L(k^0,k)[1+N(k^0)]}{k^0+\k^2/2M}\,\tau\\
-g^2\int\frac{d\k}{(2\pi)^3}\int\frac{dk^0}{2\pi}\frac{\rho_L(k^0,k)[1+N(k^0)]}{(k^0+\k^2/2M)^2}\left[1-e^{-(k^0+\k^2/2M)\tau}\right].
\end{multline}
The result, for the zero-momentum case,
is plotted in Fig.~\ref{fig:g_smalltau} and compared with the full one-loop calculation.
As it can be see, for the moderate coupling
$\alpha\!=\!0.4$ considered here, the weak coupling expansion is accurate till quite large values of $\tau/\beta$. What is perhaps surprising is the dependence on the mass $M$, which reflects a non analytic behavior at small $\tau$. Assume indeed that a Taylor expansion of Eq.~(\ref{eq:G10}) exists. Then, the leading term in this expansion, of order $\tau^2$, is obtained by setting $\tau'=\tau''=0$ in the integrand, leading to the result 
\beq
G_1(\tau)/e^{-M\tau}= \!\frac{g^2\tau^2}{2}\int\frac{d\k}{(2\pi)^3}\Delta(0,\k)=\frac{g^2\tau^2}{2}\Delta(\tau\!=\!0,\r\!=\!0),
\eeq
which would be independent of the mass $M$. 
However, as already stressed, 
$\Delta(\tau\!=\!0,\r\!=\!0)$ is divergent, so that  Eq.~(\ref{eq:G10}) has no Taylor expansion. The integral over $\tau'$ and $\tau''$ in Eq.~(\ref{eq:G10}) exists however, and because of the exponential factor, it acquires a dependence on the mass $M$: it is largest in the limit $M\to\infty$, and decreases as $M/T$ decreases. This is the trend seen in Fig.~(\ref{fig:g_smalltau}).
\section{Numerical results: MC simulations and MEM analysis}\label{sec:numerical}

In this section we present the results of the numerical evaluation of the path integral for the heavy-quark correlator. We shall also discuss the spectral density obtained from the latter through an analysis based on the Maximum Entropy Method (MEM) \cite{MEM}. Since no ambiguity can arise, we use in this section the simplified notation $G(\tau,\r)$ for the Euclidean correlator in place of $G^>(-i\tau,\mathbf
r )$ used in the rest of the paper. This correlator is obtained from the path integral derived in Sect.~\ref{sec:path_integral}.  By taking the ratio  of $G(\tau,\r)$ with the free propagator $G_0(\tau,\r)$ (see Eq.~(\ref{freesolution})) one obtains
\beq
\frac{G(\tau,\r)}{G_0(\tau,\r)}= \frac{
\int_{\0}^{\r}{\cal D}
\z\;{\rm e}^{-S_0[\z]}\,{\rm e}^{\bar F(\z)}  }
{\int_{\0}^{\r}{\cal D}
\z\;{\rm e}^{-S_0[\z]}\,
}=\langle {\rm e}{^{\bar F[\z,\tau]} } \rangle,
\label{eq:geucl}
\eeq
with  \beq 
S_0[\z,\tau]= \int_{0}^{\tau}d\tau' \,\frac{1}{2}M\dot
{\z}^2 ,
\eeq
and
\beq
\bar F[\z,\tau]=\frac{{g}^2}{2}\int_0^\tau d\tau' \int_0^\tau d\tau''
\Delta(\tau'-\tau'',\z(\tau')-\z(\tau'')).
\label{eq:geucl2}
\eeq
The functional $\bar F[\z,\tau ]$ is a known functional of the path, with $\Delta(\tau, \r)$  an intrinsic property of the plasma, calculated as indicated in Sect.~\ref{sec:model_medium}. The calculation of $G(\tau,\r)$ according to Eq.~(\ref{eq:geucl}) amounts to an average that can be performed using Monte Carlo (MC) techniques. 

\subsection{Monte Carlo evaluation of the path integral}

In fact, to proceed with the MC calculation, we shall take a slightly different route than that suggested by Eq.~(\ref{eq:geucl}). This is because we want to include the effects of the interaction in the samples of paths used in the averaging. While this may not be necessary in the present one particle problem, this is essential when dealing with the two particle problem that we plan to address in the future. Thus, using a standard strategy, we define a propagator $G_\alpha(\tau,\r)$ as in Eq.~(\ref{eq:geucl}) but with the action replaced by \beq
S_\alpha[\z,\tau]=S_0[\z,\tau]-\alpha \bar F[\z,\tau], 
\eeq 
with $\bar F[\z,\tau]$  given by Eq.~(\ref{eq:geucl2}).
Clearly, $S_\alpha[\z,\tau]$ interpolates between $S_0[\z,\tau]$, corresponding to $\alpha=0$, and the full action $S_0[\z,\tau]- \bar F[\z,\tau]$ reached for $\alpha=1$. By taking the derivative
of $\ln G_\alpha$ with respect to $\alpha$ one obtains
\begin{equation}
\frac{1}{G_\alpha(\tau,\r)}\,\frac{\partial G_\alpha(\tau,\r)}{\partial \alpha} =\frac{  \int {\cal D} \mathbf z
\;\bar F[\z]\;
 \exp\left[- S_\alpha[\mathbf z]\right]}{ \int {\cal D} \mathbf z\,\exp\left[- S_\alpha[\mathbf z]\right] }=\langle \bar F[\z] \rangle_\alpha,
\label{eq:Ga1}
\end{equation}
 and $G(\tau,\r)$ is recovered after integration over $\alpha$:
  \begin{eqnarray}
\ln \frac{G(\tau,\r)}{G_0(\tau,\r)}=\ln\left(\langle {\rm e}^{\bar F[\z,\tau]} \rangle\right) =\int_0^1 d\alpha ~ \frac{\partial \ln G_\alpha(\tau,\r)}{\partial \alpha} =
\int_0^1 d\alpha ~\langle \bar F[\z] \rangle_\alpha.
\label{eq:kirkwood1}
\end{eqnarray}
The $\alpha$-dependent average value appearing in the right-hand side of
Eq.~(\ref{eq:Ga1}) is evaluated using a MC algorithm whose details are given in Appendix~\ref{sec:details}. 
\subsection{The Euclidean correlator}
The heavy quark correlator is calculated first in coordinate
space, and then at zero spatial momentum. Calculations have been performed for a fixed mass $M\!=\!7.5$ and temperatures ranging from $T\!=\!0.75$ to $T\!=\!2$. Recall that all energies in the MC calculation are expressed in units of  $197.3$ MeV (so that $T=1$ corresponds to $T\simeq 200$ MeV, and $M=7.5$ to $M\simeq 1.5$ GeV). It is also useful to remember in the following that the ratio of propagators in the left hand side of Eq.~(\ref{eq:kirkwood1}) is a dimensionless function of $\tau/\beta, \r/\beta$, $T/M$,  and $m_D/T$. Actually, since we keep the coupling constant $\alpha\!=\!0.4$ fixed, $T/M$ is the only relevant control parameter.

\begin{figure}[!tp]
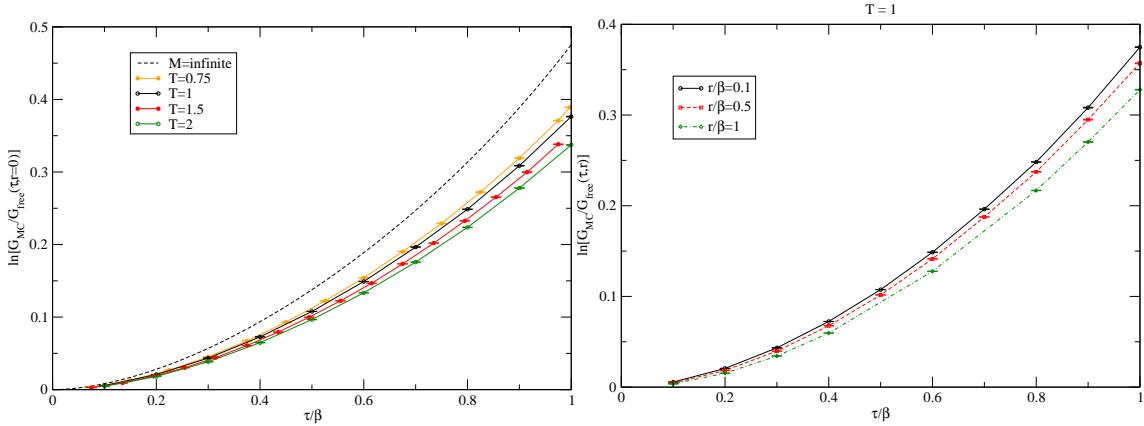

\begin{center}
\includegraphics[clip,width=0.5\textwidth]{log_Gr0.eps}\includegraphics[clip,width=0.5\textwidth]{logGtau_over_Gfree.eps}
\caption{Left panel: The quantity $\bar{F}^{\rm MC}(\tau,\r=0)$  for various temperatures (and $M\!=\!7.5$). As the ratio $T/M$ decreases the curves move closer to the static result. Right panel: $\bar{F}^{\rm MC}(\tau,\r)$  for $T\!=\!1$ and various values of $\r$.}
\label{fig:MC_coordinate}
\end{center}
\end{figure}
For the ease of presentation we define
\beq
\bar{F}^{\rm MC}(\tau,\r)\equiv\ln\frac{G(\tau,\r)}{G_0(\tau,\r)}.
\eeq
The quantity $\bar{F}^{\rm MC}(\tau,\r)$ is displayed in Fig.~\ref{fig:MC_coordinate} together with its infinite mass limit, the function $\bar F(\tau)$ (see Eq.~(\ref{eq:Ftau})). Since $\Delta(\tau,z\!=\!0)>\Delta(\tau,z)$, $\bar F[\z,\tau]<\bar F(\tau)$: hence diffusion tends to decrease the magnitude of $\bar F^{\rm MC}(\tau,\r)$. Thus, the larger the ratio $T/M$, the larger the diffusion, and the lower is the corresponding curve in the left panel of Fig.~\ref{fig:MC_coordinate}. The panel on the right hand side of Fig.~\ref{fig:MC_coordinate} indicates that the effect of the interactions depends mildly on $\r$: it attenuates very slowly as $r$ increases. 

We now consider the correlator projected to zero momentum 
\beq
G(\tau,\p=0)\equiv\int d\r\, {G}(\tau,\r).
\eeq
It is again convenient to study the ratio
\beq
e^{\bar F^{\rm MC}(\tau,\p=0)}\equiv\frac{G(\tau,\p=0)}{G_{0}(\tau,\p=0)}=\frac{\int d\r \exp[-Mr^2/2\tau]\,{G}(\tau,\r)/{G_0}(\tau,\r)}{\int d\r \exp[-Mr^2/2\tau]}.
\eeq
This expression  lends itself to a convenient numerical evaluation. Indeed, 
as an outcome of the MC simulations, for each $\tau$, one knows 
$G/G_0(\tau,\r)$ for a discrete, and rotationally symmetric, set of values $\{\r_i\}$ (typically $r_i\!\simle\!2$ fm). One can then write
\beq\label{Eq:p0recipe}
\frac{G(\tau,\p=0)}{G_{0}(\tau,\p=0)}=\frac{\sum_i r_i^2 \exp[-Mr_i^2/2\tau]\,{G}(\tau,\r_i)/{G_0}(\tau,\r_i)}{\sum_i r_i^2 \exp[-Mr_i^2/2\tau]},
\eeq 
which is the formula used to obtain $G(\tau,\p\!=\!0)$.

A further remark is in order. As explained in Appendix \ref{sec:details}, in the MC calculations, the functional $\bar F[\z]$ is truncated to the following discrete sum
\beq\label{eq:discr_sum}
\bar{F}'[\z]\equiv\frac{g^2}{2}\sum_{i\ne j=1}^{N_\tau}a_\tau^2\,\Delta((i-j)a_\tau,\z_i-\z_j).
\eeq
A procedure is introduced then to correct for the missing diagonal ($i\!=\!j$) terms, assuming that these are approximately given by the corresponding terms in the calculation of the known function $\bar F(\tau)$. This amounts to correct the raw data by the quantity
\beq\label{eq:correction}
\langle {\rm e}^{\bar F'[\z,\tau]} \rangle\to \langle {\rm e}^{\bar F'[\z,\tau]} \rangle\, {\rm e}^C,
\eeq
with 
\beq
C=\bar F(\tau)-\frac{g^2}{2}\sum_{i\ne j=1}^{N_\tau}a_\tau^2\,\Delta((i-j)a_\tau,0).
\eeq
This correction is linear in $\tau$, and affects for instance the calculation of the free energy, given by the correlator evaluated at $\tau\!=\!\beta$. Table \ref{table:correction} summarizes the results.
As we see, the correction is small and is no more than a few percent.
\begin{table}[htdp]
\begin{center}
\begin{tabular}{|c|c|c|}
\hline
T&$\bar F^{\rm MC}(\beta,\p\!=\!0)$& C\\
\hline
0.75&0.382&0.0102\\
1 & 0.366 & 0.0131\\
1.5 & 0.346 & 0.0184\\
2 & 0.331 & 0.0235\\
\hline
\end{tabular}
\caption{$\bar F^{\rm MC}(\beta,\p\!=\!0)$ for various temperatures obtained with the raw MC data and
the correction $C$ in Eq.~(\ref{eq:correction}). The values of the free-energy shift obtained here for
$T\!=\!1$ and $1.5$ can be compared with the one-loop results in Table \ref{massshift} for the
cases $T/M\!=\!0.133$ and $0.2$, respectively. In absolute values, the MC (one-loop) free energy shifts are 0.366 (0.336) and 0.346 (0.318), respectively for the two cases; the one-loop approximation underestimates the free energy shift.  }
\end{center}
\label{table:correction}
\end{table}

\begin{figure}[!tp]
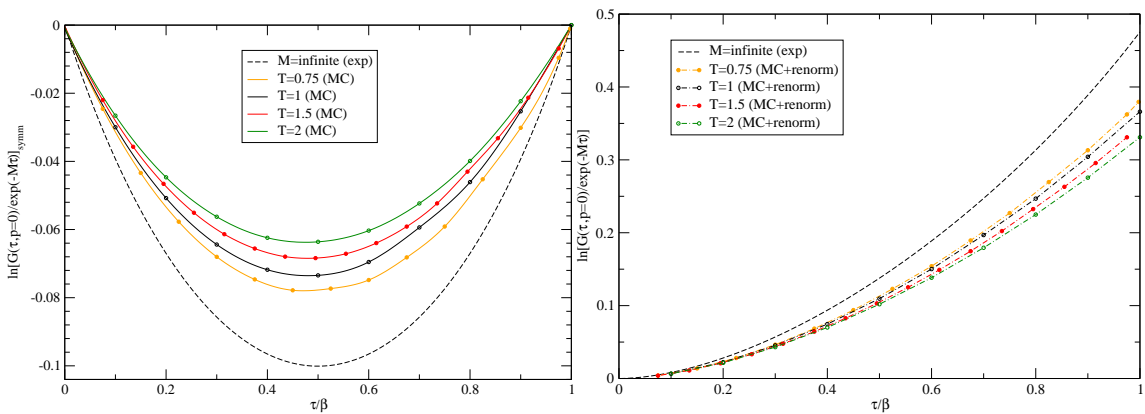

\begin{center}
\includegraphics[clip,width=0.5\textwidth]{logGp0_symm_points_new.eps}\includegraphics[clip,width=0.5\textwidth]{logGp0_over_Gfree_ren.eps}
\caption{Left panel: $\bar F_2^{\rm MC}(\tau,\p\!=\!0)$ obtained with the raw MC data, for all
the temperatures covered by our analysis. As usual $M\!=\!7.5$. Right panel: the same quantity after
the correction indicated in Eq.~(\ref{eq:correction}) (and labelles here as ``MC+renorm'').}
\label{fig:MC_Gp0/Gfree}
\end{center}
\end{figure}
The quantity $\bar F^{\rm MC}(\tau,\p\!=\!0)$ is shown in Fig.~\ref{fig:MC_Gp0/Gfree} for various
temperatures. The left panel displays $\bar F_2^{\rm MC}(\tau)\!\equiv\!\bar F^{\rm MC}(\tau)\!-\!(\tau/\beta)\bar F^{\rm MC}(\beta)$. The curves are obtained employing directly the raw MC data, which are not affected by the correction (\ref{eq:correction}). In the right panel we show the corrected results. As already found in studying the $\r\!=\!0$ correlator, the curves move closer to
the static result as the ratio $T/M$ decreases, 
due to the suppression of diffusion.

\begin{figure}[!tp]
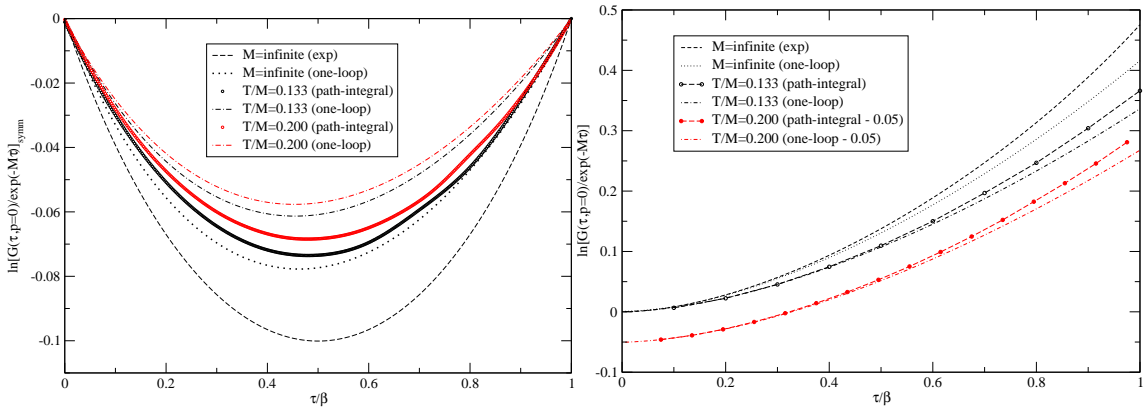

\begin{center}
\includegraphics[clip,width=0.5\textwidth]{logGp0_MCvsoneloop_symm_new.eps}\includegraphics[clip,width=0.5\textwidth]{logGp0_MCvsoneloop_new.eps}
\caption{A comparison between $\bar F^{\rm MC}(\tau,\p\!=\!0)$ and $\bar F^{(1l)}(\tau,\p\!=\!0)$ as a function of the ratio $T/M$. The one-loop curves are obtained from a numerical integration of the charm ($M=1.5$ GeV) spectral density studied in Sect.~\ref{sec:physical_processes} for $T\!=\!200$ and $T\!=\!300$ MeV. In the right panel the  set of curves corresponding to $T=300$ MeV has been translated downwards by -0.005 in order to make the figure more readable.}
\label{fig:Gp0_MCvsoneloop}
\end{center}
\end{figure}
In Fig.~\ref{fig:Gp0_MCvsoneloop} we compare the MC results with those of the one-loop calculation presented in Sect.~\ref{sec:physical_processes}, in which the Euclidean correlator was obtained through the numerical integration of the corresponding spectral function, according to Eq.~(\ref{euclcorrelspec}). The MC points start quite close to the one-loop curves corresponding to the same value of $T/M$, in agreement with the expectation that the short-time behavior is governed by perturbation theory (as already discussed in Sect.~\ref{sec:physical_processes} commenting Figs.~\ref{fig:correlator_one_loop2} and \ref{fig:g_smalltau}). For large values of $\tau/\beta$ the MC results lie above the
one-loop curves.
This general behavior is also analyzed within the simple toy-model presented in Appendix \ref{sec:toy-model}.

\subsection{The spectral function}
We turn now to the  reconstruction of the heavy quark  spectral density  from the Euclidean correlator obtained with the MC calculation. To do so, we need to invert Eq.~(\ref{euclcorrelspec}), a well known difficult problem. We use here a maximum entropy analysis (MEM), according to the algorithm described in Ref.~\cite{MEM}. Another exhaustive introduction to the method can be found in Ref.~\cite{MEM2}. In such an approach, one determines the ``best'' possible spectral function, given the information one has about the Euclidean correlator (the ``data''), and prior information one has about the spectral density, such as the fact that it is positive definite (and hence can be interpreted as a probability density) and that it satisfies some sum rules. 
The procedure involves the maximization of an entropy function (actually the minimization of a free-energy), which is defined with respect to a \emph{default model}: in the absence of data, the spectral density coming out of the entropy maximization is the default model. 
There is, of course, a delicate interplay between the effect of the data
and that of the default model 
, and the resulting  spectral density will in general  keep some reminiscence of the chosen default model. In order
to explore such a systematic uncertainty we will consider two
different default models: a constant (within the finite range $|\omega|<10$), and a Gaussian of the form
$\exp[-(\omega\!-\!M)^2/2\gamma^2]/\sqrt{2\pi\gamma^2}$.
In both cases we adjust the parameters of the default model so that the first two sum rules in Eq.~(\ref {sumrules012}) are fulfilled. 

Throughout this section the results will be expressed in terms of dimensionless variables, displaying for instance  $\sigma(\omega)T$ as a function of $\bar\omega/T=(\omega\!-\!M)/T$, the only parameter left being the ratio $T/M$.
It is then useful to recall that in the static limit $T/M\to 0$, the free-energy shift is  $-\alpha m_D/2T\!\approx\!-0.476$, while $\omega_{\rm pl}/T\!\approx\!1.373$ controls the location of the plasmon absorption/emission peaks .
 
\begin{figure}[!tp]
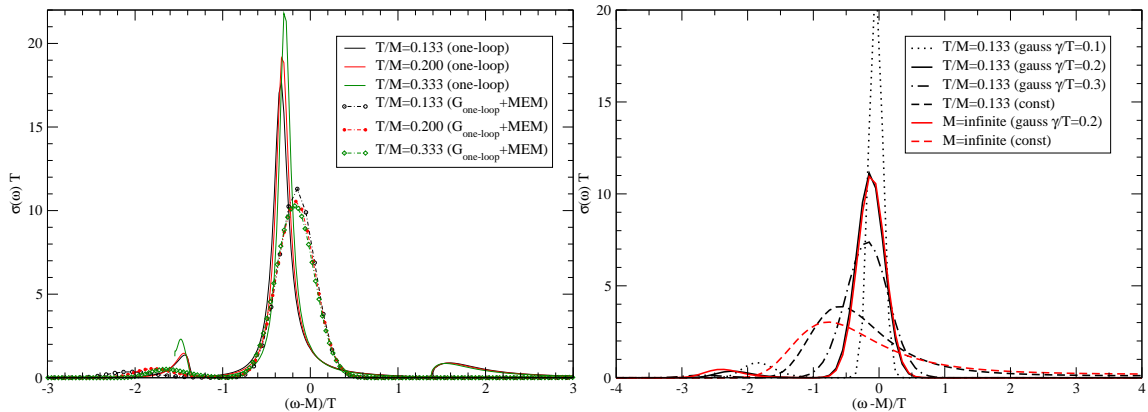

\begin{center}
\includegraphics[clip,width=0.5\textwidth]{spectral_charm_1lMEM_ToverM.eps}\includegraphics[clip,width=0.5\textwidth]{MEM_prior_ToverM.eps}
\caption{Left panel: a test of the MEM  reconstruction of the one-loop spectral function for a charm quark ($M=1.5$ GeV) at three different temperatures. A Gaussian prior is used. The shift of the main peak is systematically underestimated, and its width overestimated. Right panel: the dependence on the default model. We use a Gaussian and a constant, as explained in the text. The MEM procedure was applied to the data for $G^{\rm MC}(\tau)$ in the case $T/M\!=\!0.133$ and to the exact result for $G^{M=\infty}(\tau)$. The constant default model leads systematically to larger shift and a bigger width than the Gaussian default model.}
\label{fig:MC_sigma_system}
\end{center}
\end{figure}
As the first test of the  potentiality of the MEM procedure,   and of the systematic uncertainties attached to the choice of the default model, we reconstruct the (known) one-loop spectral density from the one-loop Euclidean correlator $G^{(1l)}(\tau)$ obtained in Sect.~\ref{sec:physical_processes},  through the integration of the corresponding spectral density.  
We use a large set of data  points ($\sim500$), and  
take a heavy quark mass $M\!=\!1.5$ GeV and  temperatures $T\!=\!200,300\;{\rm and}\;500$ MeV.
As one can see on the left panel of Fig.~\ref{fig:MC_sigma_system}, the MEM inversion -- here performed with a Gaussian prior -- is able to identify the main peak. However this is broader and less shifted then the exact result: the shift is $\sim\!-0.15$, while it is $\sim\!-0.35$ in the original one-loop spectral density.  The method also reconstructs  a low-energy secondary bump, though less pronounced than  the plasmon-absorption peak in the original one-loop spectral function, and it appears also at  lower frequency ($\sim\!-2$ compared to $-1.45$). On the other hand no signature of the high-energy secondary peak present in $\sigma^{(1l)}(\omega)$ is visible in the MEM spectral density. In the right panel of Fig.~\ref{fig:MC_sigma_system} we illustrate the sensitivity to the choice of the default model. There, the MC data at $T\!=\!1$ are used, as well as the known infinite mass correlator $G^{M=\infty}(\tau)$. We consider a Gaussian (with various values of the width) and a constant prior. With a Gaussian prior with width $\gamma/T\!=$ 0.1, 0.2 and 0.3,  the main peak is shifted respectively by, -0.05, -0.15 and -0.15.   The presence of a spectral strength at low energy seems to be a quite robust feature of the spectrum, though the broader the default model, the less pronounced the secondary bump is. In particular for a flat prior one finds simply a
very large broadening and negative shift of the main peak.  

\begin{figure}[!tp]
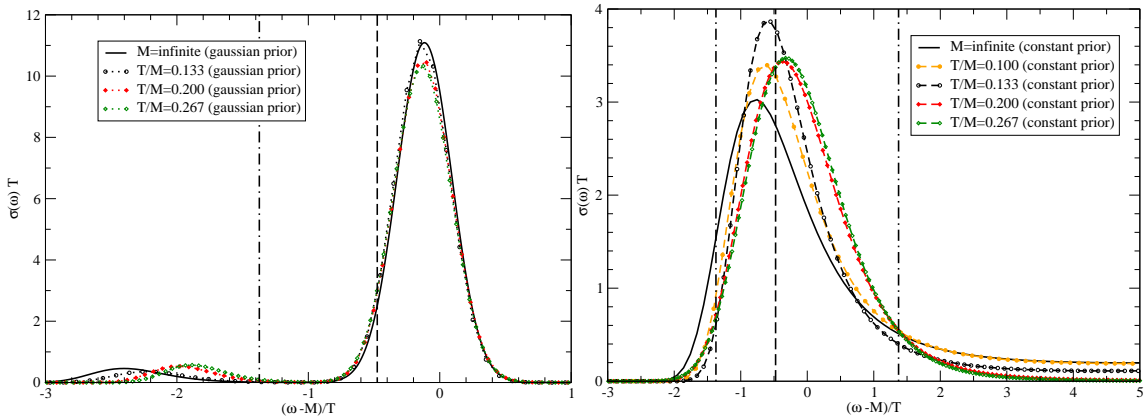

\begin{center}
\includegraphics[clip,width=0.5\textwidth]{MEM_ToverMscan_gauss.eps}\includegraphics[clip,width=0.5\textwidth]{MEM_ToverMscan_const.eps}
\caption{The MEM spectral densities $\sigma^{\rm MC}(\omega)$ for different values of $T/M$. In the left/right panel a Gaussian/constant default model is employed. For comparison the curves obtained from $G^{M=\infty}(\tau)$ are also shown. The dashed vertical lines correspond to the static free-energy shift $-\alpha m_D/2T=0.476$. The dot-dashed vertical lines signal $\pm\omega_{\rm pl}$, where $\omega_{\rm pl}/T=1.373$ is the plasma frequency. As clearly seen in the left panel, the Gaussian prior leads to an underestimate of the shift of the main peak (here estimated as the static free energy shift), together with an overestimate of that of the secondary peak at low energy (here estimated by $-\omega_{pl}$). The dependence on $T/M$  is very weak. On the right panel one sees that the dependence on $T/M$ is larger with the constant prior, and in line with what one expects (the curves move gradually towards that corresponding to the infinite mass limit as $T/M$ decreases).}
\label{fig:MC_sigma}
\end{center}
\end{figure}
In Fig.~\ref{fig:MC_sigma} we show the results of the MEM inversion of the MC data for $G^{\rm MC}(\tau)$, for various values of $T/M$, including the exact infinite mass limit corresponding to $T/M\!=\!0$.
The left panel corresponds to a Gaussian default model with $\gamma\!=\!\alpha T/2$. The resulting spectral
densities present a broad main peak, slightly shifted with respect to its position in the vacuum ($M$) by an
amount roughly proportional to $T$ (the curves in the dimensionless units employed lie almost on top of each others), but smaller (by a factor $\sim\!5$) than the static free-energy shift $-\alpha m_D/2T\!\approx\!0.476$. A secondary low-energy bump, more and more displaced with respect to the main peak as $T/M$ decreases, is also visible.
In the right panel the same data are analyzed using  a constant default model. In such a case the spectral function exhibits only a broad peak with  a sizable negative shift which, as $T/M\!\to\!0$,
results $\sim\!50\%$ larger than the static free-energy shift.
Furthermore the MEM spectral density, with this choice for the prior, has also a long high-energy tail, at variance with what is  found with the Gaussian default model.

\begin{figure}[!tp]
\begin{center}
\includegraphics[clip,width=0.8\textwidth]{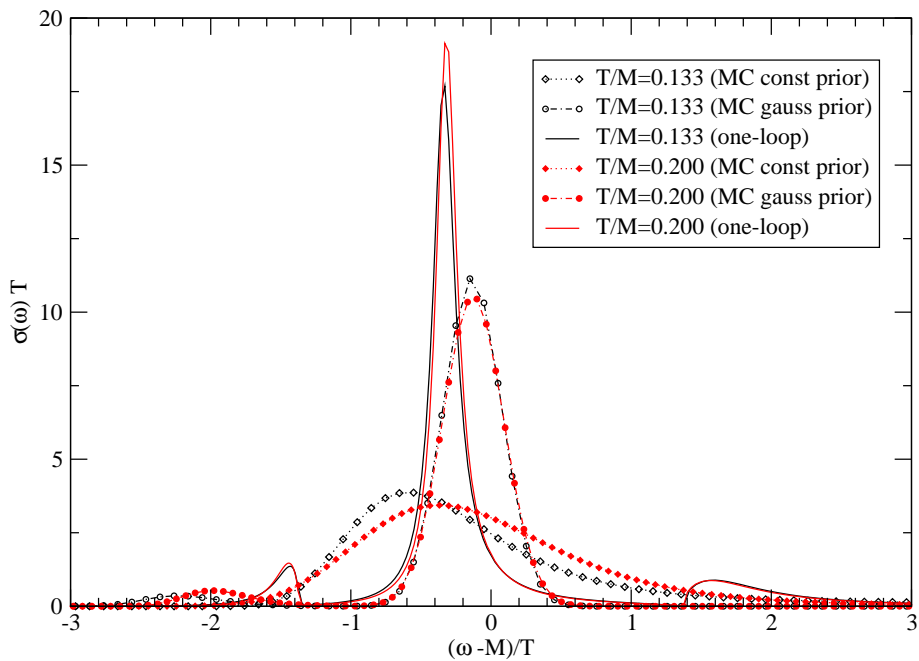}
\caption{The MEM spectral densities $\sigma^{\rm MC}(\omega)$ obtained with the two different default models (dotted and dot-dashed curves), compared to $\sigma^{(1l)}(\omega)$ (continuous curves), for two values of $T/M$.}
\label{fig:MEVvsoneloop}
\end{center}
\end{figure}
Finally in Fig.~\ref{fig:MEVvsoneloop} a comparison between the MEM and the one-loop spectral functions is given. The main features discussed above can be seen. In particular,  the dependence on the default model is striking. A Gaussian leads to a very small shift of the main peak. On the other hand the constant default model yields a broader and more shifted peak, whose strength extends to low energy till displaying a
partial overlap with the secondary bump of the curve obtained with the Gaussian prior. 

\section{Conclusions}\label{sec:conclusions}

In this paper, we have presented an approach to the dynamics of heavy quarks in a hot plasma based on a path integral for non relativistic particles with a non local (in space and time) self-interaction that summarizes the effects of the medium to which the heavy quark is coupled. 

The path integral providing the heavy-quark Euclidean correlator was evaluated numerically using Monte Carlo techniques. The results of this numerical evaluation were analyzed and compared to those of the one-loop calculation, and to those of an exact evaluation of the path integral in the infinite mass limit. We showed that the effect of interactions is to favor the contribution of straight paths in the path integral, and are indeed  maximum in the infinite mass limit, where the heavy particle stays at rest. 
Calculations were done for a value of the coupling constant that would correspond in QCD to a value of the strong coupling constant $\alpha_s\approx .3$. For such a value the one-loop approximation provides a reasonable first approximation, but deviations with the exact Monte Carlo  results were obtained. The Monte Carlo results move towards  those of the infinite mass limit as the ratio  $M/T$ increases, as expected. 

The Monte Carlo
calculations of the Euclidean correlator were performed in coordinate space, but  a simple integration over the spatial coordinates gave the correlator for 
zero-momentum. This allowed us, in particular, to estimate the shift in the  free energy of the system that is caused by the addition of the heavy quark. We also used the corresponding Euclidean correlator to reconstruct the   spectral
function, through a MEM analysis. Within our implementation of this method, we were able only to reproduce the main qualitative features, namely a broad main peak, whose shift is only given semi-quantitatively. A secondary structure below the main peak, somewhat reminiscent of the plasmon-absorption peak of the one-loop spectral function is also seen, but no secondary structure above the main peak is detected, only a long tail at large frequencies is observed (and only with a constant prior).
The large sensitivity of the MEM analysis to our choices of default models does not allow us  to draw more
robust quantitative conclusions at this stage. On the other hand, the qualitative features that we were able to reconstruct may be enough to draw conclusions in the two particle problem, which is our ultimate goal. 

 The thorough analysis of the one-particle case that we have presented in this paper paves the way for several extensions. Clearly the calculation can be improved in several places, and the general setting brought closer to QCD without too much efforts. For instance, we have seen that the  HTL approximation used in the description of the hot plasma leads to a somewhat unrealistic description of the effects of collisions. While this affects only mildly the heavy quark correlator, and only at small times where the calculation is in complete control (being essentially perturbation theory), this feature can be improved without changing the basic structure of the problem. 
The calculation of the Euclidean correlator of a heavy quark-antiquark pair is within reach. The reconstruction of the spectral density of a heavy quark pair  from its Euclidean correlator faces the same difficulty as met in lattice QCD: on the one hand, this offers opportunities for more detailed comparisons between the two approaches, on the other hand we  note that our path integral for the Euclidean correlator can be calculated with  high precision, which could be exploited to develop new methods of reconstruction of the spectral density. Finally one may contemplate the possibility of calculating the path integral directly in real time, perhaps at the cost of additional approximations. That would allow us to bypass the problem of the analytical continuation, and would open the possibility of numerous applications.  

\section*{Acknowledgments}
A.B. and J.P.B. gratefully acknowledge ECT* for warm hospitality and financial support during the preparation of this work.
\appendix%

\section{An exactly solvable toy-model}\label{sec:toy-model}

In this section we present a toy model  illustrating some of the features of the  calculations that are presented in the main text, in particular those features that emerge in the infinite mass limit. 
The model consists of  a fermion of mass $M$ coupled to a single harmonic oscillator that represents the ``medium''. The Hamiltonian of the system is written as
\beq
H=M\psi^\dagger\psi+
\frac{1}{2}\,\left(\pi^2+m_D^2\, \phi^2\right)
+g\,\psi^\dagger\psi\,\phi,\qquad \phi\equiv \frac{a+a^\dagger}{\sqrt{2m_D}},
\eeq
where $\psi^\dagger$ and $\psi$ are the creation and the annihilation operators of the fermion, $\{\psi,\psi^\dagger\}=1$, $\phi$ and $\pi$ are respectively  the coordinate of the oscillator and its conjugate momentum, $[\phi, \pi]=i$, and 
$a^\dagger, a$ the associated  creation and annihilation operators, $[a,a^\dagger]=1$. Since $[H,\psi^\dagger\psi]=0$, the eigenstates of  $H$ can be classified in sectors characterized by the eigenvalue   of the fermion number operator $\psi^\dagger\psi$. Since the fermion has no internal degree of freedom there are only two sectors to consider, those with $\psi^\dagger\psi=0$ and with $\psi^\dagger\psi=1$. The first sector  corresponds to the medium without the fermion, and  the Hamiltonian is simply that of the oscillator
\beq
H_0=m_D\,\left(a^\dagger\,a+1/2\right).
\eeq
The sector with $\psi^\dagger\psi=1$ mimics the case in which one adds the  fermion  into the medium.  The corresponding  Hamiltonian reads
\beq
H_1=M+H_0+ g\phi\equiv M+H_0+V,
\eeq 
and it has the structure of Eq.~(\ref{eq:hamiltoniandecomp}). It is easily diagonalized by introducing
the shifted operators
\beq
b\!\equiv\! a\!+\!\frac{g}{\sqrt{2m_D^3}}\quad{\rm and}\quad
b^\dagger\!\equiv\! a^\dagger\!+\!\frac{g}{\sqrt{2m_D^3}},\quad{\rm with}\quad[b,b^\dagger]=1,
\eeq
so that
\beq
H_1=\left(M-\frac{g^2}{2m_D^2}\right)+m_D\left(b^\dagger\,b+\frac{1}{2}\right).
\eeq
The spectrum of $H_1$ is identical to that of $H_0$, and the shift in the  ground-state energy is given by:
\beq\label{energyshift}
\Delta{\cal E}\equiv {\cal E}_1-{\cal E}_0=M-\frac{g^2}{2m_D^2}\equiv M-\alpha m_D,\qquad \alpha\equiv \frac{g^2}{2m_D^3},
\eeq
where we have introduced the dimensionless coupling constant $\alpha$. 
The  ground state of $H_1$ is a coherent state characterized by a non-vanishing expectation value 
of the field $\phi$:  
\beq
\langle \phi\rangle= -\frac{g}{m_D^2}=-\sqrt{\alpha}\sqrt{\frac{2}{m_D}}.\label{phieq}
\eeq
This expectation value plays the role of the classical field $A_0$ associated with the polarization cloud around the heavy quark.

One may also consider the non-equilibrium situation that corresponds to adding the  fermion  into the system in its ground state at $t=0$. Following this initial perturbation,  the whole system evolves in time with the Hamiltonian $H_1$. It is then not difficult to establish that the expectation value of $\phi$ oscillates  around its equilibrium value (\ref{phieq}) according to 
\beq
\langle\phi\rangle_t=\langle\phi\rangle_{eq} (\cos m_D t\, -1),
\eeq
where $\phi_{eq}$ is given by Eq.~(\ref{phieq}).

This result holds unchanged when the oscillator is in thermal equilibrium at temperature $T$, that is, $\langle\phi\rangle$ is not affected by thermal fluctuations. Similarly, because the spectra of $H_1$ and $H_0$ are identical, the contributions of thermal fluctuations cancel in the difference of free energies of the systems with and without the  fermion, with the result that this  difference remains equal to the shift in the ground state energy given by Eq.~(\ref{energyshift}). 

Consider now the Euclidean correlator
\beq
G(\tau)\equiv G^>(-i\tau )\equiv\langle\psi(\tau)\psi^\dagger(0)\rangle_0,
\eeq
where the expectation value $\langle\dots\rangle_0\!\equiv\!{\rm Tr}\left[e^{-\beta H_0}\dots\right]/Z_0$ is taken over states of the medium without the  fermion.
One has:
\beq
G^>(-i\tau)=\langle e^{H\tau}\psi e^{-H\tau}\psi^\dagger\rangle_0=
\langle e^{H_0\tau}\psi e^{-H_1\tau}\psi^\dagger\rangle_0=
e^{-M\tau}\left\langle e^{H_0\tau}e^{-(H_0+V)\tau}\right\rangle_0,
\eeq
where, in the last expression,
one recognizes the evolution operator in the interaction representation, so that one can write:
\beq
G^>(-i\tau)=e^{-M\tau}\left\langle T_\tau\exp\left[-g\int_0^\tau d\tau'\phi_I(\tau')\right]\right\rangle_0.
\eeq  
A simple calculation yields then the exact result:
\beq\label{defF0}
G^>(-i\tau)={\rm }e^{-M\tau}\,{\rm }e^{\bar F(\tau)}\,, \eeq
where 
\beq\label{defF}
\bar F(\tau)=\frac{g^2}{2}\int_0^\tau d\tau'\int_0^\tau d\tau''\Delta(\tau'-\tau'').
\eeq
Here  $\Delta(\tau)$ is the Euclidean propagator for the field $\phi$, satisfying periodic boundary conditions ($\Delta(0)=\Delta(\beta)$): 
\beqa\label{propaphi}
\Delta(\tau)&=&\langle T\phi_I(\tau)\phi_I(0)\rangle=\frac{1}{2m_D}\left[  {\rm e}^{-m_D|\tau|} (1+N)+{\rm e}^{m_D|\tau|} N  \right],
\eeqa
with $N$ the statistical factor
\beq
N\equiv \frac{1}{{\rm e}^{\beta m_D}-1}.
\eeq

At this point let us note that the model depends on several dimensionful parameters: the mass $M$, which simply shifts the overall spectrum, and plays no role in the dynamics; the Debye mass $m_D$ which characterizes the response of the system to an external perturbation, such as the addition of the  fermion;  the coupling constant $g$ and the temperature $T$. We shall systematically express the coupling between the fermion  and the oscillator  in terms of the dimensionless coupling $\alpha$, as in Eq.~(\ref{energyshift}). A look at the propagator (\ref{propaphi}) reveals that $m_D^{-1}$ appears there as the natural time scale, while the statistical factor depends on $m_D/T$.

It is sometimes convenient to write $\bar F(\tau) $ as the sum of a term $\bar F_1(\tau)$ linear in $\tau$ and a term $\bar F_2(\tau)$ that is symmetric around $\beta/2$:
\beqa\label{defF1F2}
\bar F_1(\tau)&=& \alpha m_D \tau ,\nonumber\\
\bar F_2(\tau)&=& \alpha \left[   \frac{  \cosh (m_D(\tau-\beta/2))-\cosh(\beta m_D/2)}{\sinh\beta ( m_D/2)  }  \right].
\eeqa
Clearly, 
\beq\label{barFDelta}
\bar F(\beta)=\bar F_1(\beta)=\alpha m_D\beta=\frac{g^2}{2}\beta \Delta(i\omega_n=0),
\eeq
so that $M-(1/\beta)\bar F_1(\beta)=\Delta F_Q$ is the difference of free energies of the systems with and without the  fermion (see Eq.~(\ref{energyshift})). The last equality in Eq.~\ref{barFDelta} emphasizes that $\bar F(\beta)$ is entirely given by the zero Matsubara frequency part of the oscillator propagator (\ref{propaphi}), with 
\beq
\Delta(i\omega_n)=\int_0^\beta d\tau {\rm  e}^{i\omega_n\tau}\Delta(\tau).
\eeq
The function $\bar F_2(\tau)$ vanishes at $\tau=0$ and $\tau=\beta$, by construction,  and has its minimum at $\tau=\beta/2$, with  value $\bar F_2(\beta/2)=-\alpha \tanh (\beta m_D/4)$. The slope at $\tau=0$ is $-\alpha m_D$, so that the  linear contributions cancel in $\bar F=\bar F_1+\bar F_2$, in accordance with the general result (see Eq.~(\ref{der1t0})).  This is also obvious from Eq.~(\ref {defF}): the small $\tau$ behavior starts at order $\tau^2$.  At quadratic order, we have
\beq
\bar F(\tau\ll m_D^{-1})\simeq \frac{g^2\tau^2}{2}\langle \phi^2\rangle=   \frac{1}{2} \alpha m_D^2 \tau^2 (1+2N), \qquad \langle \phi^2\rangle=\frac{1}{2m_D}(1+2N)=\Delta(\tau=0).
\eeq
One recovers the general result between the coefficient of $\tau^2$ and the fluctuation of $\phi$ (see Eq.~(\ref{der2t0})). We shall return to the short time behavior of the correlator shortly. 

We now   exploit the analyticity of $G^>$ and move to real time. This will allow us in particular to get the large time behavior of $G^>(t)$. One gets from Eq.~(\ref{defF0})
\beq\label{realtG}
G^>(t)={\rm e}^{-iMt}\, {\rm e}^{iF(t)}, 
\eeq
with  
\beq
F(t)= \frac{g^2}{2}\int_0^t ds\int_0^t ds' D(s-s'),\quad D(s-s')=i\Delta(\tau=is, \tau'=is').
\eeq
A simple calculation then yields
\beq
F(t)= g^2 \int_{-\infty}^{\infty} \frac{d\omega}{2\pi}\frac{1-\cos\omega t}{\omega^2} D(\omega),
\eeq
with $D(\omega)$ the Fourier transform of the time-ordered propagator $D(t)$ (see Eq.~(\ref{eq:realtFourier})). The large time behavior of the correlator  follows then from  Eq.~(\ref{eq:limcos}):
\beq
F(t\gg m_D^{-1})=\frac{g^2}{2} t D(\omega=0)= \alpha m_D t.
\eeq
It is entirely determined by the static response of the medium.
The comparison with Eq.~(\ref{energyshift}) reveals that $-F(t)/t$ is the interaction contribution to the energy shift caused by the addition of the fermion (see also Eq.~(\ref{barFDelta})). 
This is similar to what happens in the case of an infinitely massive quark although, in the latter case, a damping term also appears next to the free energy shift. No such term appears here because of the discrete nature of the spectrum.

One can also calculate the spectral function. To do so, it is convenient
to start with the following explicit expression of the propagator (\ref{realtG}):
\begin{multline}
G^>(t)=\exp\left[-\alpha(1\!+\!2N)\right]\exp\left[-i(M\!-\!\alpha m_D)t\right]\times\\
\times\exp\left[\alpha \left(N e^{im_Dt}+(1\!+\!N)e^{-im_Dt}\right)\right],
\end{multline}
and expand the last exponential in powers of $\alpha$.  One gets
\beq
G^>(t)={\rm e}^{-\alpha(1\!+\!2N)}{\rm e}^{-i(M\!-\!\alpha m_D)t}
\sum_{n=0}^{\infty}\frac{\alpha^n}{n!}\sum_{p=0}^{n}\binom{n}{p}(N)^p e^{ip m_Dt }(1\!+\!N)^{n-p}e^{-i\,(n-p)m_D t}.
\eeq
The Fourier transform is then obtained immediately and reads
\beqa
\sigma(\bar \omega)=2\pi\,{\rm e}^{-\alpha(1\!+\!2N)}
\sum_{n=0}^{\infty}\frac{\alpha^n}{n!}\sum_{p=0}^{n}\binom{n}{p}(N)^p(1\!+\!N)^{n-p}\delta\left(\bar\omega+\alpha m_D-(n\!-\!2p)m_D\right),\nonumber\\ \label{eq:spectral}
\eeqa
where we have set $\bar\omega\equiv \omega-M$.
The above spectral density exhibits an infinite number of peaks in one-to-one correspondence with the transitions between the eigenstates of $H_1$. The major peak is located at $\bar\omega=-\alpha m_D$. 
The expansion of the spectral density to order $\alpha^K$  has peaks centered at $\bar\omega=-\alpha m_D \pm k m_D$, with $k=0,1\dots K$. Hence, the larger the coupling, the larger the number of peaks giving a sizable contribution to the spectral density. 

\begin{figure}[!tp]
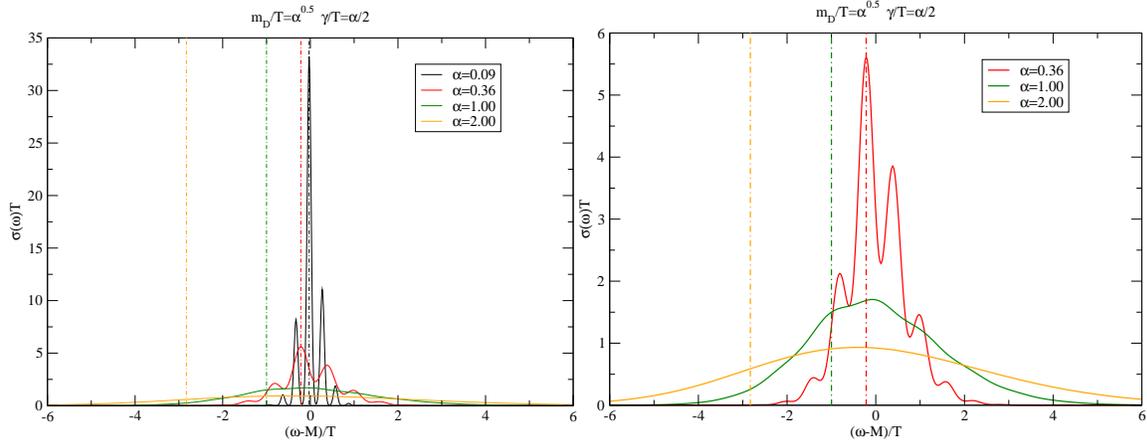

\begin{center}
\includegraphics[clip,width=0.50\textwidth]{def_spectral.eps}\includegraphics[clip,width=0.50\textwidth]{def_spectral_zoom.eps}
\caption{The spectral function (\ref{eq:spectral}), for different values of the coupling $\alpha$. The vertical lines refer to
the position of the ``main peak'' at $\omega\!=\!M\!-\alpha m_D$. The delta functions have been smeared to gaussians with width $\gamma=\alpha T/2$, and the mass $m_D$ is adjusted as a function of the temperature, $m_D= T\sqrt{\alpha}$. As the coupling grows, the individual peaks are smoothed out, leaving a broad, structureless, distribution.}
\label{fig:spectral_exact} 
\end{center}
\end{figure}
 The  spectral density is displayed  in Fig.~\ref{fig:spectral_exact}   for a wide range of values of the coupling $\alpha$.  In order to make contact with the general discussion of a heavy quark in a plasma, we choose $m_D=\sqrt{\alpha}T$ (this implies among other things that the coupling among the plasma grows similarly  the coupling between the fermion and the plasma particles).   Also, for the purpose of illustrating the global behavior of the spectral function, we smear the  delta functions by replacing them by gaussians of variance $\gamma\!\sim\!\alpha T$. At small coupling, individual peaks are recognized. For large coupling, the smearing that we have introduced erases the individual secondary peaks, leaving a broad distribution which spreads over a larger and larger frequency interval as the coupling grows. Note that the main peak, located at $\bar\omega=-\alpha m_D$ is shifted to lower frequency as $\alpha $ grows, but the spectral strength remains centered around $\omega\sim M$.  This behavior may be understood in terms of the sum rules satisfied by the spectral function. 

These sum rules are obtained from the derivatives s of $G^>(t)$ at $t=0$:
\beq
i^n \left. \frac{\partial^n }{\partial t^n}{\rm e}^{iMt}G^>(t)\right|_{t=0}=\int_{-\infty}^{\infty} \frac{d\bar\omega}{2\pi}\,  \bar\omega^n \sigma(\bar\omega).
\eeq
These derivatives are easiy calculated and one gets
\beqa\label{sumrules0123}
\int_{-\infty}^{\infty} \frac{d\bar\omega}{2\pi}\,   \sigma(\bar\omega)=1, \qquad \int_{-\infty}^{\infty} \frac{d\bar\omega}{2\pi}\,  \bar\omega\,\sigma(\bar\omega)=0,\qquad \qquad\qquad\qquad\nonumber\\  \int_{-\infty}^{\infty} \frac{d\bar\omega}{2\pi}\,  \bar\omega^2\, \sigma(\bar\omega)=\alpha m_D^2(1+2N),\quad \int_{-\infty}^{\infty} \frac{d\bar\omega}{2\pi}\,  \bar\omega^3\, \sigma(\bar\omega)=\alpha m_D^3.
\eeqa
These sum rules explain why the spectral weight remains centered around $\bar \omega=0$, with a width increasing with $\alpha$, while the last sum rule suggest a somewhat larger strength at positive $\bar \omega$ than at negative $\bar \omega$. 
Note that the sum rules that are displayed explicitly here  are at most linear in the coupling $\alpha$. The first higher order correction, of order $\alpha^2$, enters at the level of the $\omega^4$ sum rule.

\begin{figure}[!tp]
\begin{center}
\includegraphics[clip,width=0.60\textwidth]{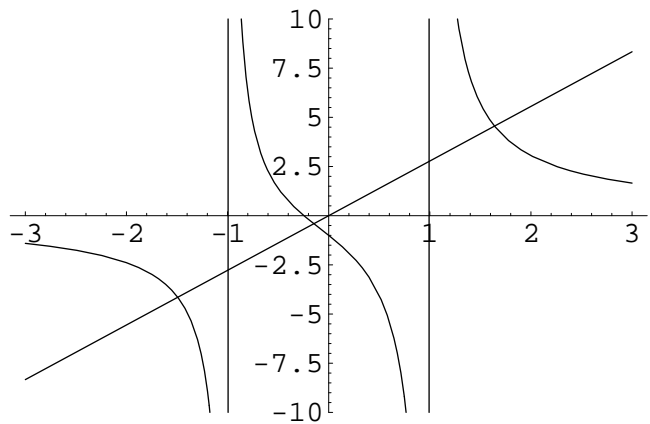}
\caption{Graphical solution of the equation $\bar\omega=\Sigma(\bar\omega)$, with both $\Sigma$ and $\bar\omega$ expressed in units of $m_D$ (the curves plotted are $\Sigma(\bar\omega/\alpha$ and $\bar\omega/\alpha$). The self-energy (divided by $\alpha$) exhibits poles at $\bar\omega=\pm m_D$. The intersections with the straight line $\bar\omega/\alpha$  give the locations of the poles of the propagator. There is always one pole close to $\bar\omega=0$. At weak coupling this pole has the largest residue (the straight line in the figure corresponds to $\alpha=0.36$). When the coupling grows the other two poles move away as $\approx \pm \alpha m_D$, and their residue saturate the sum rule, leaving very little spectral weight on the pole at $\bar\omega\approx 0$ (which asymptotically moves to $\bar\omega=-m_D/(1+2N)$). Note that the intersection of $\Sigma$ with the vertical axis yields the exact energy shift, $\Sigma(\bar\omega=0)=-\alpha m_D$. }
\label{fig:sigmaTM} 
\end{center}
\end{figure}

Let us now turn to the one-loop approximation for the time-ordered (or retarded) propagator. The one-loop self-energy of the fermion is easily obtained:
\beq
\Sigma(\bar\omega+i\eta)=\alpha m_D^2\left[ \frac{1+N}{\bar\omega-m_D+i\eta} +\frac{N}{\bar\omega+m_D+i\eta} \right].
\eeq
The poles of $\Sigma$ for $\bar\omega=\pm m_D$ correspond to the energies of the fermion having emitted or absorbed a quantum of the oscillator, which represent the leading processes that take place at weak coupling. The inverse retarded propagator reads
\beq
G^{-1}(\bar\omega+i\eta)=-\bar\omega-i\eta+\Sigma(\bar\omega+i\eta).
\eeq
Thus, the propagator has three poles, at values $\bar\omega_i$ solutions of the equation
\beq\label{poleeqn}
\bar\omega^3-\bar\omega \,m_D^2[1+\alpha(1+2N)]-\alpha m_D^3=0.
\eeq
The general behavior of the solutions may be easily inferred from the graph displayed in Fig.~\ref{fig:sigmaTM}. 
The propagator may then be written as 
\beq
G(\bar\omega)=\sum_i \frac{z_i}{\bar\omega_i-\bar\omega},
\eeq
with the residues given by
\beq
z_i^{-1}=1-\partial \Sigma/\partial \bar\omega|_i.
\eeq
 The spectral function takes the form
\beq\label{eq:spectal1l_toy}
\sigma(\bar\omega)=2\pi \sum_i z_i \,\delta(\bar\omega-\bar\omega_i). 
\eeq 
It can be verified that, in the weak coupling limit, this coincides with the general expression (\ref{eq:spectral}) expanded to order $\alpha$. The Euclidean correlator is easily obtained from the spectral function, and reads
\beq
G^>(-i\tau)={\rm e}^{-M\tau}\,\sum_i z_i \, {\rm e}^{-\bar\omega_i \tau}={\rm e}^{-M\tau}{\rm e}^{\bar F^{(1l)}(\tau)},
\eeq
which defines the function $\bar F^{(1l)}(\tau)$.
From the correlator calculated for $\tau=\beta$, one deduces the one-loop free energy shift 
\beq
\bar F^{(1l)}(\beta)=\ln\left[ \sum_i z_i \,{\rm e}^{-\bar\omega_i \beta}\right].
\eeq
This is to be compared to the exact value $\bar F(\beta)=\alpha m_D/T$: the one-loop calculation 
underestimates the exact result.

\begin{figure}[!tp]
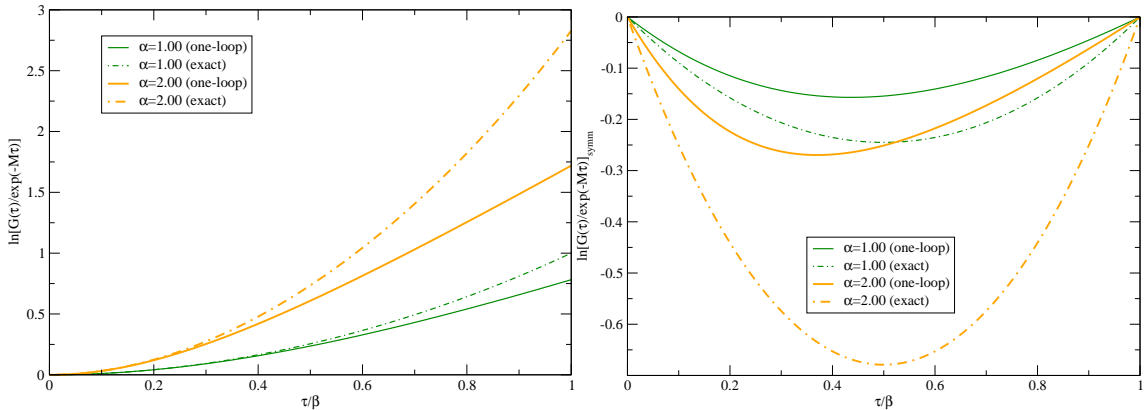

\begin{center}
\includegraphics[clip,width=0.50\textwidth]{gtau_toy_log_strong.eps}\includegraphics[clip,width=0.50\textwidth]{gtau_toy_symm_strong.eps}
\caption{Left: The function $\bar F^{(1l)}(\tau)$ (continuous curves) compared to the exact result $\bar F(\tau)$ (dot-dashed curves) as a function of $\tau/\beta$ for large values of the coupling constant $\alpha\!=\!1$ and $\alpha\!=\!2$ . Even in this strong coupling regime, the one-loop approximation gives an excellent approximation to the exact result up to values $\tau\simle \beta/2\alpha$. Right: The same for $\bar F_2^{(1l)}(\tau)$ and $\bar F_2(\tau)$. The slope at the origin is a measure of the free energy shift, which can be also read on the left panel as the value of $\bar F(\beta)$. Both plots exhibits clearly that the free-energy shift is underestimated in the one-loop approximation. Note also the asymmetry, growing with increasing coupling, of the one-loop results with respect to $\tau =\beta/2$, in contrast to the exact curves.}
\label{fig:correlator1} 
\end{center}
\end{figure}
A comparison between the one-loop and the exact result is offered in Fig.~\ref{fig:correlator1}. 
As we did earlier, we may decompose $\bar F^{(1l)}(\tau)=\bar F^{(1l)}_1(\tau)+\bar F^{(1l)}_2(\tau)$, with $\bar F^{(1l)}_1(\tau)=(\tau/\beta) \bar F^{(1l)}(\beta)$. The function $\bar F^{(1l)}_2(\tau)$ is plotted in the right panel of Fig.~\ref{fig:correlator1}.
The agreement of the exact and one-loop correlators may be understood from the fact that the  sum rules (\ref{sumrules0123}) are exactly satisfied at one loop, namely
\beq\label{sr0123oneloop}
\sum_i z_i=1,\quad \sum_i z_i \bar\omega_i=0,\quad  \sum_i z_i \bar\omega_i^2=\alpha m_D^2 (1+2N) ,\quad  \sum_i z_i \bar\omega_i^3=\alpha m_D^3. 
\eeq
To these we should add the relation $\sum_i \bar\omega_i=0$, that derives immediately  from Eq.~(\ref{poleeqn}). The sum rules (\ref{sr0123oneloop}) are the one-loop transcription of the exact sum rules mentioned above, Eq.~(\ref{sumrules0123}). They hold exactly at one-loop because the small time behavior of the propagator involves also a small $g$ expansion and, as it has already been mentioned after Eq.~(\ref{sumrules0123}), up to order $\tau^3$, the small $\tau$ expansion involves terms of the weak coupling expansion only  up to order $g^2$. Such terms are taken into account exactly by the one-loop self energy. The fact that
the one-loop result is sufficient to describe the small-$\tau$ behavior appears clearly in the left
panel of Fig.~\ref{fig:correlator1} where, even for large values of the coupling, the one-loop and exact correlators are hardly distinguishable  for $\tau/\beta$ small enough.

\begin{figure}[!tp]
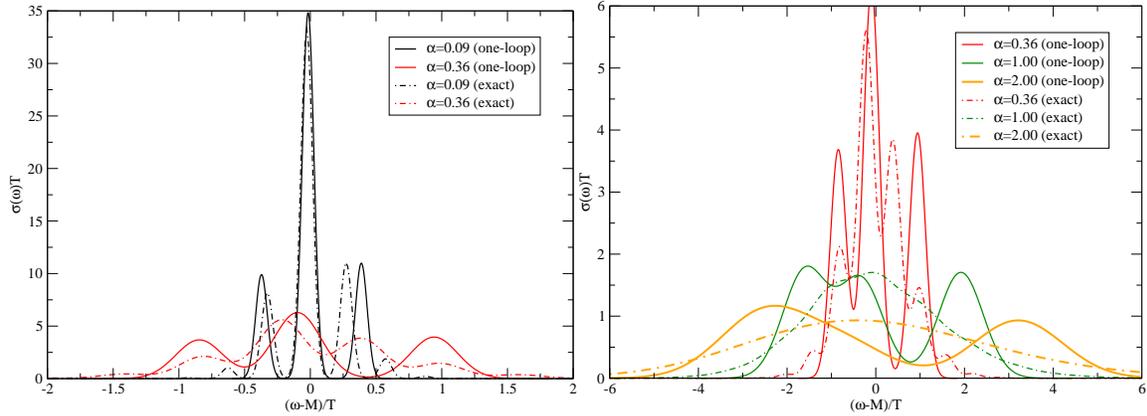

\begin{center}
\includegraphics[clip,width=0.50\textwidth]{spectral_toy_weak.eps}\includegraphics[clip,width=0.50\textwidth]{spectral_toy_strong.eps}
\caption{The one-loop spectral function compared to the exact one. The curves refer to different values of the coupling $\alpha$, from weak (left panel) to strong (right panel). The one-loop result (continuous curves), characterized by the presence of only three peaks, is compared to the exact one (dot-dashed curves), which has a richer structure. In plotting the curves the delta functions have been smeared to gaussians in the same way as in Fig.~\ref{fig:spectral_exact}.}
\label{fig:spectral_toy} 
\end{center}
\end{figure}
Finally in Fig.~\ref{fig:spectral_toy} we provide a comparison between the one-loop spectral function,
given by Eq.~(\ref{eq:spectal1l_toy}), and the exact result. For weak coupling they look quite similar.
On the other hand for larger coupling more and more secondary peaks contribute to the exact spectrum,
while the one-loop result can display only three peaks. These, having to fulfill the sum-rules
(\ref{sr0123oneloop}), result largely distorted.

\section{Details on the path integral implementation}\label{sec:details}

The path integral that we want to evaluate has the following form (see Eq.~(\ref{eq:Ga1})):
\beq
\langle \bar F[\z,\tau]\rangle_\alpha=\frac{\int_0^{\r} {\cal D}\z \, \bar F[\z,\tau] \, {\rm e}^{-S_\alpha[\z,\tau]}}{\int_0^{\r} {\cal D}\z \, {\rm e}^{-S_\alpha[\z,\tau]}}.
\eeq
For any chosen value of $\tau=N_\tau a_\tau$, where $N_\tau$ is an integer and $a_\tau$ a fixed time interval,  the paths  are defined by a discrete set of points $\{\z(\tau_i)\}$, where $\tau_i$ ($0\le \tau_i\le \tau$) is a multiple of $a_\tau$. We choose natural units $\hbar = c = k_B = 1$ and fix the
unit of length to  be 1~fm, and correspondingly  the unit of
energy (or temperature) to be $197.3$~MeV. The time step is fixed at the value $a_\tau=0.01$ fm/c, and the heavy quark mass at $M=7.5$ (corresponding approximately to the mass of a charm quark).

The path integral Monte Carlo method is based on the generation of a Markov
chain  that 
samples a set of paths according to the distribution 
\begin{equation}
W_\alpha[\mathbf z] = \frac{\exp(-S_\alpha[\z])}{\int [{\cal D}\z] \, {\rm e}^{-S_\alpha[\z]}}
.
\end{equation}
Then the average $\langle \bar F[\z] \rangle_\alpha$ is evaluated as the
arithmetic average over the paths generated by the 
equilibrated Markov chain.

We used the Metropolis algorithm to generate the Markov chain by a sequence of elementary moves. A move is defined as follows:  starting from a given path $\z$, one selects at random a time $\tau_i$, and 
displace the corresponding point $\z(\tau_i)$ by a quantity $\delta \z$ uniformly distributed in a cube of side $d$ centered at $\z(\tau_i)$, thus defining a new path $\z'$;  the move is accepted with the probability
\begin{equation}
\pi = \min \left[1,\frac{\exp(-S_\alpha[\z'])}{\exp(-S_\alpha[\z])}\right].
\end{equation}
 In the present calculation we start from a straight path connecting
the origin $(0,0)$ to the point $(\tau,\r)$, and perform at least
$10^5 \times N_\tau$ moves to reach equilibrium. During this stage the
value of $d$ is adjusted to keep the acceptance ratio of attempted
moves between $0.45$ and $0.55$. Typical values of $d$ were found in
the range $0.07-0.08$, at the temperature $T = 1.0$. Once the Markov chain has reached equilibrium, one continues generating paths, and the corresponding paths are used in calculating  the average values of $\langle \bar F[\z] \rangle_\alpha$. At least $10^5 \times N_\tau$ paths of the equilibrated chain  are used in the calculation of the average value. Finally, the integrand appearing in the right hand side of Eq.~(\ref{eq:kirkwood1}) is evaluated on an equispaced array of $10$ points in the interval from $\alpha =
0$ to $\alpha = 1$. The resulting curve is then interpolated with a cubic
spline and integrated using the adaptive Gauss--Kronrod method as
implemented in the GNU Scientific Library~\cite{GSL}.

At a given temperature we take tipically between 10 and 20 values of $\tau$ to determine $G(\tau,r)$, with $r$ varying between 0 and 2 fm.
Because a large number of paths are used in the calculation of the average, the statistical errors are negligible: one gets relative errors of order $10^{-6}$ for small $\tau$ and $10^{-4}$ for the largest values of $\tau$.  Furthermore, since  the average  is taken over a different set of trajectories at each $\tau$, the results at various $\tau$ are uncorrelated.


There is one issue in this calculation that deserves further comments. It concerns the calculation of the integral (\ref{eq:geucl2}). The simplest discretized form of this integral reads 
\beq\label{eq:discr_sum2}
\bar{F}[\z,N_\tau]\equiv\frac{g^2}{2}\sum_{i, j=1}^{N_\tau}a_\tau^2\,\Delta((i-j)a_\tau,\z_i-\z_j).
\eeq
This, however, cannot be used as it stands since, as we have seen in Sect.~\ref{sec:model_medium}, $\Delta(0,0)$ is logarithmically divergent, so that the diagonal terms $i=j$ in the expression above are ill defined. Before we explain how we have gone around this difficulty, let us examine the calculation of the same integral in the infinite mass limit, where the paths are frozen at the origin, i.e., $\z(\tau_i)=0$ for all $i$. Then the functional $\bar F[\z,\tau]$ reduces to  the function  $\bar F(\tau)$ (Eq.~(\ref{eq:Ftau})), that we may write, using the same discretization as above but for the ``diagonal'' terms, as
\beq\label{eq:discr_sum3}
\bar{F}(\tau)\approx\frac{g^2}{2}\sum_{i\ne  j=1}^{N_\tau}a_\tau^2\,\Delta((i-j)a_\tau,0)+N_\tau \bar{F}(a_\tau),
\eeq 
where $\bar F(a_\tau)$ is the exact value of the integral on a square of side $a_\tau$. The comparison of the value of $\bar F(\tau)-N\bar F(a_\tau)$ calculated exactly, and from the discretized sum in Eq.~(\ref{eq:discr_sum3}), yields an estimate of the discretization error in the evaluation of the integral. As can be seen in  Fig.~\ref{fig:F(a)}, this error is of (relative) order $10^{-3}$ and increases slightly towards small values of $\tau$. We note also that for $\tau/\beta\simge 0.2$, the contribution of the diagonal terms amounts to less than 10\%. We have exploited these features in order to make the following simplifications:
\begin{figure}[!tp]
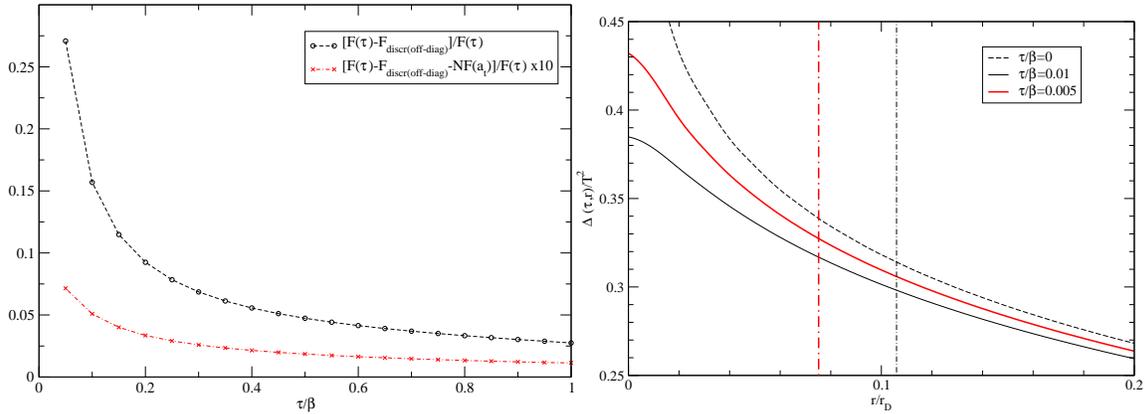

\begin{center}
\includegraphics[clip,width=0.5\textwidth]{ratio_offdiag.eps}\includegraphics[clip,width=0.5\textwidth]{deltaLmD_short.eps}
\caption{Left panel: the relative error in evaluating $\bar{F}(\tau)$ with the discretization algorithm employed for the path integral ($N_\beta=100$). The black dashed curve is obtained by simply dropping the $i\!=\!j$ terms in the sum, as in Eq.~(\ref{eq:discr_sum4}); the red dot-dashed curve arises after correcting with the term $N_\tau \bar F(a_\tau)$ from Eq.~(\ref{eq:discr_sum3}).  Right panel: The short-time/distance behavior of the HTL propagator $\Delta(\tau,\r)$, for various values of $\tau/\beta$ (blow-up version of Fig.~\ref{fig:Delta}). The vertical lines refer to the corresponding values of $\,\overline{r}\!\equiv\!\sqrt{\langle r^2(\tau)\rangle}$.}\label{fig:F(a)}.
\end{center}
\end{figure}

i) Only the off-diagonal terms are used in the sampling of paths, that is $\bar F[\bar z,\tau]$ is replaced for that purpose by 
\beq\label{eq:discr_sum4}
\bar{F'}[\z,N_\tau]\equiv\frac{g^2}{2}\sum_{i\ne  j=1}^{N_\tau}a_\tau^2\,\Delta((i-j)a_\tau,\z_i-\z_j).
\eeq

ii) A correction is applied to compensate for the omission of the diagonal terms, assuming this correction to be given by $N_\tau \bar F(a_\tau)$ (in practice we calculate this correction from the difference between $\bar F(\tau)$ and $\bar F'[\z,\tau]$ in Eq.~(\ref{eq:discr_sum4}) estimated  for $\z=0$). Note that this correction is presumably an overestimates. Indeed because of diffusion, at time $\sim a_\tau$, the heavy quark is on the average at a distance $\bar r=\sqrt{3\tau/2M}$ away from the origin, and 
$\Delta({\tau},\bar r)<\Delta({\tau},0)$. A quantitative measure of this overestimate (which is of the order 30\%) can be read off the right panel of Fig.~\ref{fig:F(a)}.

A final source of errors comes from the fact that the MC calculation has been set in fixed physical units. Thus, as we change the temperature, of equivalently $\beta=1/T$, one varies the number of discretization points, with $T=1$ ($\sim 200$ MeV) corresponding to $N_\beta=100$. Increasing $T$, means decreasing $\beta$, and correspondingly $N_\beta$. Because of this, the calculations become less  precise as the temperature increases. 


\end{document}